%% file: main.tex
\documentclass[
    9pt,
    twocolumn,
    ]{extarticle}

\usepackage{amsmath}
\usepackage{amssymb}
\usepackage[utf8]{inputenc}

\usepackage[
    colorlinks=true,
    linkcolor=black,
    citecolor=black,
    filecolor=black,
    urlcolor=blue,
    plainpages=false,
    pdfpagelabels,
    breaklinks=true,
    pdfusetitle,
    ]{hyperref}

\usepackage[
    natbib=true,
    backend=biber,
    citestyle=numeric-comp,
    sorting=none,
    giveninits=true,
    uniquename=init, 
    url=false
    ]{biblatex}
\usepackage{bm}
\usepackage{booktabs}
\usepackage[font=small,labelfont=bf]{caption}
\usepackage{changes}
    \definechangesauthor[color=red]{MAP}
    
\usepackage[capitalise, nameinlink]{cleveref}
    \crefname{section}{Sec.}{Secs.}
\usepackage[sharp]{easylist}
    \ListProperties(
        Hang=true,Space=4pt,Space*=4pt,Hide=1000,
        Margin1=1.5em,Style1*=\textbullet\hskip .5em,
        Margin2=3.7em,Style2*=--\hskip .5em,
        Margin3=5.9em,Style3*=$\ast$\hskip .5em,
        Margin4=7.8em,Style4*=$\cdot$\hskip .5em
        )
\usepackage[a4paper, margin=20pt]{geometry}
\usepackage[docdef=atom]{glossaries-extra}
    \setabbreviationstyle[acronym]{long-short}
    \glssetcategoryattribute{acronym}{nohyperfirst}{true}
    
\usepackage{graphicx}
\usepackage{lipsum}
\usepackage{listings} 
\usepackage{multirow}
\usepackage{parskip}
\usepackage{placeins}
\usepackage{setspace}
\usepackage{subfigure}
\usepackage{siunitx}
\usepackage{threeparttable}
\usepackage{todonotes}

\usepackage{xspace}

\newcommand{\den}{\text{den}}
\newcommand{\exc}{\text{exc}}
\newcommand{\inh}{\text{inh}}

\newcommand{\leak}{\text{l}}

\newcommand{\som}{\text{som}}
\newcommand{\syn}{\text{syn}}
\newcommand{\tb}[1]{\textbf{#1}}
\newcommand{\tgt}{\text{tgt}}

\newcommand{\nin}{n_\text{in}}
\newcommand{\nout}{n_\text{out}}

\usepackage{changes}
\definechangesauthor[color=teal]{BvH}
\definechangesauthor[color=olive]{LK}

\makeatletter

\makeatletter

\input{glossary}
\title{ELiSe: Efficient Learning of Sequences in Structured Recurrent Networks}

\newcommand{\affilDP}{\textsuperscript{1}}
\newcommand{\affilINI}{\textsuperscript{2}}
\newcommand{\affilCL}{\textsuperscript{3}}
\newcommand{\correspondingAuthor}{$^{\dagger}$}

\author{
    \normalsize{
        Laura Kriener$^{*,}$\affilDP$^{,}$\affilINI,
        Ben von Hünerbein$^{*,}$\correspondingAuthor$^{,}$\affilDP,
        Kristin Spicher\affilDP$^{,}$\affilCL,
      }\\
    \normalsize{
        Federico Benitez\affilDP,
        Timo Gierlich\affilDP,
        Arno Granier\affilDP,} \\
    \normalsize{
        Walter Senn\affilDP,
        Mihai A. Petrovici\correspondingAuthor$^{,}$\affilDP}\\
    \footnotesize{
        $^{*}$ Shared first authorship
        \hspace{3em}
        }\\
    \footnotesize{
        \correspondingAuthor Corresponding authors \{ben.vonhuenerbein, mihai.petrovici\}@unibe.ch)
        }\\[-3pt]
    \footnotesize{
        \affilDP
        Department of Physiology, University of Bern, 3012 Bern, Switzerland.
        }\\[-3pt]
    \footnotesize{
        \affilINI
        Institute of Neuroinformatics, University of Zurich and ETH Zurich, 8057 Zurich, Switzerland.
        }\\[-3pt]
    \footnotesize{
        \affilCL
        Catlab Engineering GmbH, 82284 Grafrath, Germany.
        }
}

\begin{document}

\maketitle

\input{sections/abstract}

\input{sections/introduction}

\section{Model description}\label{sec:model}

\input{sections/development}

\input{sections/learning}

\input{sections/results}

\input{sections/discussion}

\FloatBarrier

\input{sections/methods}

\printbibliography
\addcontentsline{toc}{section}{References}

\subsection*{Acknowledgments}\label{sec:ack}
\input{sections/ack.tex}

\subsection*{Author contributions}\label{sec:authcont}
\input{sections/authcont.tex}

\subsection*{Code availability}
\textbf{Code for the simulations is available at \url{https://github.com/unibe-cns/ELiSe}.}

\subsection*{Competing Interests statement}
The authors declare no competing interests.

\clearpage
\onecolumn
\input{sections/supplement}

\end{document}

%% file: glossary.tex
\newacronym[plural=ANNs,firstplural=artificial neural networks (ANNs)]{ann}{ANN}{artificial neural network}
\newacronym{bptt}{BPTT}{backpropagation through time}
    \newcommand{\bptt}{\gls{bptt}\xspace}
\newacronym{elise}{ELiSe}{Efficient Learning of Sequences}
    \newcommand{\elise}{\gls{elise}\xspace}
\newacronym{htm}{HTM}{hierarchical temporal memory}
    \newcommand{\htm}{\gls{htm}\xspace}
\newacronym{hvc}{HVC}{HVC} 
    
\newacronym{lman}{LMAN}{lateral magnocellular nucleus of the anterior nidopallium}
    \newcommand{\lman}{\gls{lman}\xspace}
\newacronym{mse}{MSE}{mean squared error}
    
\newacronym{ppc}{PPC}{posterior parietal cortex}
    
\newacronym{ra}{RA}{robust nucleus of the archistriatum}
    \newcommand{\ra}{\gls{ra}\xspace}
\newacronym{rflo}{RFLO}{random feedback online learning}
    \newcommand{\rflo}{\gls{rflo}\xspace}
\newacronym{rtrl}{RTRL}{real-time recurrent learning}
    \newcommand{\rtrl}{\gls{rtrl}\xspace}
\newglossaryentry{follow}{name=FOLLOW, description={FOLLOW Algorithm}}
    \newcommand{\follow}{\gls{follow}\xspace}
\newglossaryentry{force}{name=FORCE, description={FORCE Algorithm}}
    \newcommand{\force}{\gls{force}\xspace}

%% file: sections/abstract.tex
\begin{abstract}
    Behavior can be described as a temporal sequence of actions driven by neural activity.
    To learn complex sequential patterns in neural networks, memories of past activities need to persist on significantly longer timescales than the relaxation times of single-neuron activity.
    While recurrent networks can produce such long transients, training these networks is a challenge.
    Learning via error propagation confers models such as FORCE, RTRL or BPTT a significant functional advantage, but at the expense of biological plausibility.
    While reservoir computing circumvents this issue by learning only the readout weights, it does not scale well with problem complexity.
    We propose that two prominent structural features of cortical networks can alleviate these issues: the presence of a certain network scaffold at the onset of learning and the existence of dendritic compartments for enhancing neuronal information storage and computation.
    Our resulting model for Efficient Learning of Sequences (ELiSe) builds on these features to acquire and replay complex non-Markovian spatio-temporal patterns using only local, always-on and phase-free synaptic plasticity.
    We showcase the capabilities of ELiSe in a mock-up of birdsong learning, and demonstrate its flexibility with respect to parametrization, as well as its robustness to external disturbances.
\end{abstract}

%% file: sections/introduction.tex
\section{Introduction}

Despite seeming mundane, everyday activities such as catching a ball, walking down a flight of stairs or whistling a tune are comprised of complex and precisely timed sequences of actions.
These action sequences are grounded in corresponding spatio-temporal patterns of neural activity in motor and pre-motor areas.
One prominent example of sequence learning and replay is song production in male zebra finches, who learn to reproduce the stereotypical vocalizations of their elders.
After learning, the neural activity in the \ra pre-motor area is marked by sparse bursting that is closely linked to features in the song \cite{hahnloser2002ultrasparse, leonardo2005ensemble}.
A similarly action-coupled temporally precise activity has been observed following motor learning in mouse primary motor cortex \cite{peters2014emergence, adler2019somatostatinexpressing} and navigation learning in rat posterior parietal cortex \cite{harvey2012choicespecific}.

A detailed model of these mechanisms at the computational level remains an open question.
A key challenge for models of sequence learning is to bridge the temporal gap between neural activity evolving at the timescale of milliseconds and behaviour performed at the timescale of seconds.
Typically, sustained dynamics in recurrently connected neuron populations are assumed to produce the required transients.
The most straightforward way to exploit such activity is to learn a direct mapping to a readout population, as proposed by reservoir computing \cite{lukosevicius2009reservoir}.
However, such reservoirs are rather sensitive to initial conditions, and incorrect initialization can lead to exploding or fading activity.
Furthermore, this approach scales poorly to more complex tasks, as the readout must rely on useful transients being available within the otherwise random activity of the reservoir.
It thus seems essential for stability and resource efficiency to solve the challenging problem of learning the recurrent weights themselves.

Multiple approaches for recurrent weight learning have been previously explored.
Techniques from Machine Learning are not straightforward to apply, as they are often at odds with observed neurophysiology.
Such constraints exclude otherwise powerful methods such as \bptt \cite{werbos1990backpropagation} and \rtrl \cite{zipser1989subgrouping}.
Approximations of \rtrl such as \rflo \cite{murray2019local} improve its biological plausibility by reducing its memory requirements at the cost of approximating the propagated gradients.
However, some locality issues remain: synaptic eligibility traces need to share time constants with afferent neurons, and correct transport of errors requires copying synaptic weights \cite{murray2019local}.
Other approaches further narrow the biological plausibility gap at the expense of requiring meticulously specified forms of error transport.
Still, several important issues remain to be addressed.
FORCE learning \cite{sussillo2009generating,nicola2017supervised} requires strong feedback connections and near-instantaneous plasticity capable of closely tracking any changes in neuronal activity.
FOLLOW \cite{gilra2017predicting} partly alleviates this problem, but still relies on global error feedback to each neuron in the network.

A different approach \cite{maes2020learning} suggests the prior learning of a ``clock'' - a chain of recurrent modules that are activated in sequence.
In a second phase, arbitrary sequences can then be learned as a simple one-to-one mapping from individual links in this chain to specific motor outputs at every point in time.
While biologically plausible, this model requires large amounts of resources, as the clock chain must scale with the pattern length and complexity.
Nonetheless, the effectiveness of this model indicates that emulating innate structures can be advantageous for sequence learning, as opposed to learning that starts from a blank  slate.

Indeed, there is overwhelming evidence for such a balance happening in biological brains, between the flexibility offered by experience-dependent plasticity and the effectiveness of hardwired structural adaptations to specific use cases \citep{zador2019critique}.
The building of such scaffolding structures has been observed during development of several subsystems of the mammalian brain, such as  cerebellum, hippocampus and cortex \citep{kano2018multiple,cossart2022development,chang2021development}, as we discuss in more detail below.

\begin{figure*}[t]
    \centering
    \includegraphics[width=0.75\textwidth]{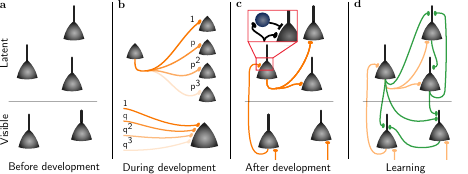}
    \caption{\tb{Development of somatic scaffold for the learning of dendritic connections.}
        \tb{(a)} The network composed of structured pyramidal neurons is divided into a visible and latent population.
        \tb{(b)} During development, axons (orange) extend to form a sparse scaffold of soma targeting connections.
                 Its structure is controlled by parameters $p$ and $q$.
        \tb{(c)} The connections of the somatic scaffold are static and have associated delays.
                 Inhibition in the scaffold is mediated by interneurons.
        \tb{(d)} Following development, dendrite targeting synapses (green) are learned according to a three-factor plasticity rule.
    }\label{fig:development}
\end{figure*}

Introducing biologically inspired components can often feel like constraining a computational model's generality instead of aiding its functionality.
In contrast, in our model of \elise we show that biasing our model in two specific ways not only increases its biological plausibility but also serves to enhance its functionality.
First, whereas deep learning builds on the training of randomly initialized neural networks -- what can properly be called a \emph{tabula rasa} -- learning in the brain builds on pre-existing structures, shaped by eons of evolution and cast into genes.
Here we suggest how such structures could form during early development, before any learning takes place, and how they can provide a better basis for recurrent learning and significantly ease the task of learning temporal sequences.
Second, neurons in deep learning models, unlike cortical neurons, do not have an internal structure and serve as simple
nonlinear transfer functions.
However, both theoretical and experimental studies have shown that the structured nature of biological neurons allows for much more complex information storage and computation within single neurons \cite{larkum2013cellular, stuart2016dendrites, london2005dendritic, poirazi2020illuminating, payeur2019classes, asabuki2022neural}.

Inspired by these experimental results, we use a two-compartment extension to simple point neuron models, in conjunction with a local error-correcting plasticity rule.
This alleviates the necessity for non-local quantities or global error signals in the learning mechanism.
Taken together, the introduction of structure both in the network and in the model neurons themselves creates a versatile substrate for continuous, on-line sequence learning in the brain.
Importantly, our model makes efficient use of its neuronal resources, allowing the learning of complex sequences with only a small number of neurons.
We demonstrate these features in a mock-up of birdsong learning, in which our networks learn long, non-Markovian\footnote{We use "non-Markovian" as shorthand for "not 1st-order Markovian".}
sequences (e.g. a sample of Beethoven's ``F\"ur Elise'') that they can reproduce robustly despite severe external disturbances.

%% file: sections/development.tex
In evolved biological systems there is a close connection between function and form.
When considering complex behavior, even if learning would start from some kind of \emph{tabula rasa}, realistically it needs the right kind of \emph{tabula} to be effective.
In order to model this, it is essential to consider early brain development.
For example, recent research \citep{cossart2022development} has shown that the functional organization of the adult hippocampus is not only formed through experience-dependent plasticity but partly hardwired at the earliest stages of development, including embryonic neurogenesis.
These developmental programs provide a scaffold onto which later experience of the external world can be grafted.
This is reﬂected in the dynamics of the adult CA1 region of hippocampus, which operates through a combination of plastic and rigid cells, bound together within segregated functional assemblies that support stable internal dynamics \citep{grosmark2016diversity,malvache2016awake}.
Similarly, Purkinje cells (PC) are thought to constitute a scaffold for sequence learning in the cerebellum. Multiple developmental phases of climbing fiber synapse elimination targeting PC cells have been observed to build such a scaffold \citep{kano2018multiple} .
Similar phases of escalating development have been observed in the auditory cortex \citep{chang2021development}.
Conversely, these scaffolds are generally pruned in the final parts of development, yielding the emphasis to experience-dependent plasticity. This is clearly observed in cortex \citep{chklovskii2004cortical,kalisman2005neocortical,lebe2006spontaneous}, where after initial development, synaptic formation, elimination, and modification are triggered by reciprocal neuronal activity. Initial geometrical connectivity therefore provides the potential for reconfigurations of local microcircuits. 

These biological observations suggest a simplified model consisting of two phases of development and learning, with development creating a scaffold structure that is exploited and modified during learning (see \cref{fig:development} and discussion below).
In the specific case of sequence learning, the neural network that emerges after the developmental process must support the stable representation of structured sequences of activity.
Further, the network should make optimal use of available neuronal resources while the connectivity matrix should remain sparse to prevent exploding activity.

\subsection{Early development provides a scaffold for learning}

As an idealization of how such a network structure could emerge, we consider a stochastic model of cortical development for two neuronal populations of pyramidal neurons
, a ``visible''~population used for explicit representation of activity sequences and a ``latent''~population for storage and recall (\cref{fig:development}a).
These populations can be considered as corresponding to pre-motor (latent) and motor (visible) areas.
Thus, our model first mimics a developmental phase that results in a scaffold of nudging connections, which is later used for transporting target signals to the latent population during learning.

During this phase, soma targeting connections are formed stochastically (\cref{fig:development}b).
The visible population of neurons is assumed to all have a somatic afferent to receive a nudging signal for learning.
Starting with the visible population, each neuron that receives at least one somatic afferent forms a number $\nout$ of efferent soma targeting connections with decreasing probability $p^{\nout}$ (for details, see \cref{sec:meth-dev}).
These connections are then accepted by postsynaptic latent neuron partners with a decreasing probability $q^{\nin}$, where $\nin$ represents the number of afferents they already have.
During this phase, all neurons (from both visible and latent populations) only project into the latent population.
With appropriate choices of parameters $p$ and $q$ for the generating functions, this yields a homogeneously and sparsely connected network, where $p$ controls the sparsity and $q$ the uniformity.
To model both excitation and inhibition, each soma targeting connection is considered to consist of a direct excitatory connection between the two neurons and additional inhibitory pathway via an interneuron (see \cref{fig:development}c and \cref{sec:meth-syn}).
While we model all soma targeting connections to be of equal strength, signals transmitted by different connections incur different, randomly drawn, transmission delays.
Including delays emulates the characteristics of activity at the neuronal and sub-neuronal (axonal and dendritic) level, known to introduce temporal delays  \citep{sreenivasan2019delay,bakkum2013tracking,zylberberg2017mechanisms}. 
Similar treatments of delays are common in the modeling literature \citep{izhikevich2006polychronization}.
This network constitutes our starting point for learning, and can be seen as the scaffold which transports the teaching signal via ``nudging'', i.e., transmitting target signals throughout the network that allow local plasticity to learn complex temporal sequences.

For modeling dendritic development, after the initial scaffolding phase, we set up weak
all-to-all connections between somata and dendritic compartments of all neurons in both populations (\cref{fig:development}d).
During learning, only dendrite targeting synapses are plastic; the soma targeting scaffold remains fixed after the initial developmental phase.

\begin{figure*}[t]
    \centering
    \includegraphics[width=1.00\textwidth]{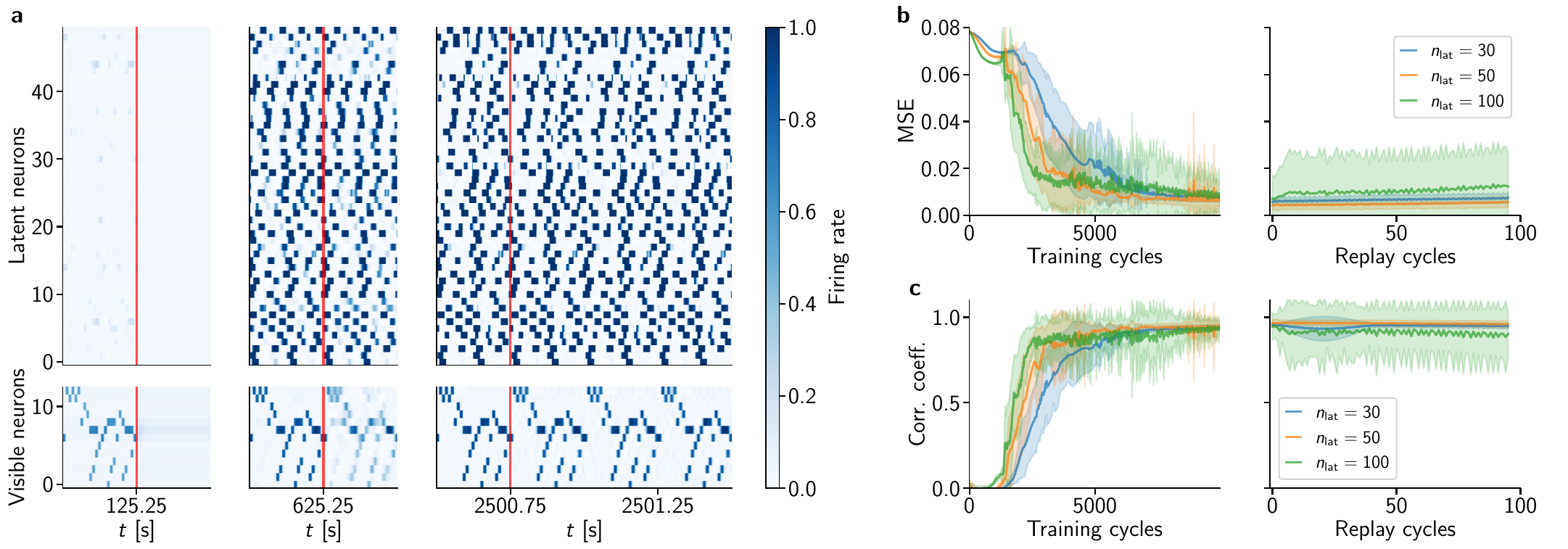}
    \caption{\tb{Learning process and replay ability of the network.}
        \tb{(a)} Evolution of network activity during learning.
        Visible neurons are first nudged towards a particular target and then released in order to observe the network’s spontaneous activity (as demarcated by the red line).
        Snapshots during early (left), intermediate (middle) and final (right) stages of training.
        \tb{(b)}
        Performance of networks with different latent population sizes measured by the \gls{mse} loss between the activity of the visible neurons and their target activity.
        Top: Evolution of the error during validation intervals, i.e.\ first replay of the pattern after being released from external nudging) during training.
        Bottom: Stability of replay after training, for continuous observation over many replay cycles following release from nudging.
        \tb{(c)} Same as (b) but using the cross-correlation between visible and target activities as a performance measure.
    }\label{fig:function}
\end{figure*}

%% file: sections/learning.tex
\subsection{Neuronal dynamics and dendritic learning}\label{sec:model_description}

\begin{figure*}[t]
    \centering
    \includegraphics[width=1.0\textwidth]{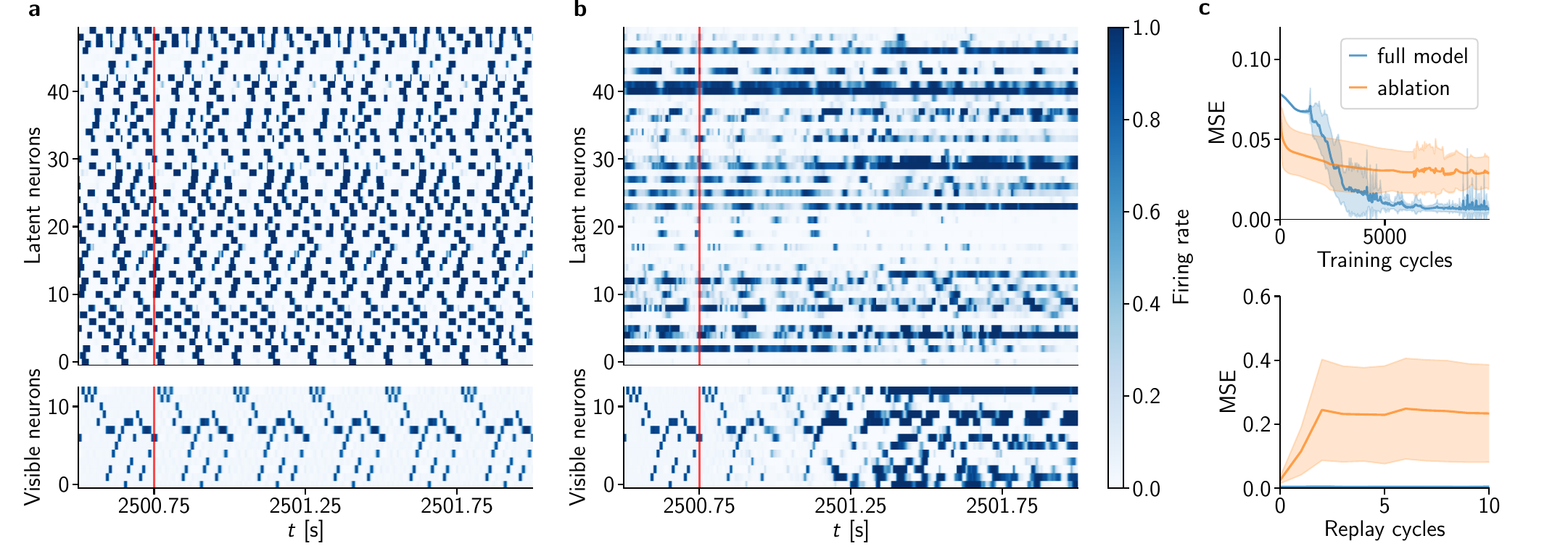}
    \caption{\tb{Comparison of replay performance between networks with and without learning of latent connections.}
        \tb{(a)} Recorded activity of the visible (bottom) and latent (top) neurons of the fully plastic network during free replay of the pattern after training.
        The red line marks the end of the external input from the teacher.
        \tb{(b)} Same as (b) but for the ablated network.
        \tb{(c)} Accuracy and stability of sequence replay, measured by the \gls{mse} between generated and target activity.
        Top: \Gls{mse} between target and produced pattern during training.
        Bottom: \Gls{mse} during continued replay after training.
    }
    \label{fig:reservoirComparison}
\end{figure*}

We consider here a rate-based model in which both the somatic and the dendritic compartment are taken to be leaky integrators
\begin{align}
    C^{\den}_\text{m}\dot v &= - g_\leak \left(v - E_\leak\right) + I^\den_\syn \\
    C^\som_\text{m} \dot u &= - g_\leak \left(u - E_\leak\right) + g_\den \left(v - u\right) + I^\som_\syn
\end{align}
with $v$ as the dendritic voltage, $u$ the somatic voltage, $C_\text{m}$ the compartmental capacitances, $g_\leak$ the leak conductance, $E_\leak$ the resting potential and $g_\den$ the coupling conductance between the dendritic and somatic compartments.

$I^\den_\syn$ and $I^\som_\syn$ represent the synaptic currents arriving at the dendritic compartment via the dendrite targeting connections and at the somatic compartment via the soma targeting connections, respectively.
The somatic compartment integrates the teaching input received via nudging and the input from the dendritic compartment.
Synaptic input onto the dendritic compartment represents the driving force of neuronal activity and is modeled as current-based, to effectively account for signal propagation along the dendritic tree.
Direct synaptic input to the soma is comparatively weak and conductance-based, thus only nudging neuronal activity into a particular direction.
This type of neuronal compartmentalization, along with the interplay between driving and nudging synapses, has been shown to enhance the computational and learning capabilities of single neurons and the networks they form \cite{urbanczik2014learning, poirazi2020illuminating, london2005dendritic, sacramento2018dendritic}.
For simplicity, we model the interaction between soma and dendritic tree as unidirectional (see also \cite{urbanczik2014learning}).
The firing rate $r$ of a neuron is modeled as a logistic function of the somatic potential $r = \varphi\left(u\right)$, which also normalizes firing rates to the interval $[0, 1]$.
For more details on the neuron and synapse models, we refer to \cref{sec:meth-nrn,sec:meth-syn}.

In addition to providing a more realistic model for the spatial structure of biological neurons, our multi-compartmental setup also allows for the use of a local error-correcting learning rule (similar to \citep{urbanczik2014learning}) for the dendrite targeting weights from neuron $i$ to neuron $j$
\begin{align}
    \dot w^\den_{ji} = \eta \, \left[\varphi\left(u_j\right) - \varphi\left(v^*_j\right)\right] \bar r_i \label{plasticity}
\end{align}
with a learning rate $\eta$ and $\bar r$ the low-pass filtered presynaptic rate.
The weight updates minimize the difference between the somatic fining rate $\varphi(u)$, which is nudged towards a target behaviour by the soma targeting connections, and the dendritic prediction of somatic firing $\varphi(v^*)$ (for additional details see \cref{sec:meth-plast}). Learning modifies the synaptic weights of these connections, of which typically only a small proportion of stronger synapses become dominant, emulating dendritic pruning in biological brains.
With this, the dendrite learns to drive the somatic voltage to produce the desired behavior as defined by the soma targeting nudging.
Learning converges once the latent population correctly drives the visible neurons to reproduce the teacher activity. At this point the teaching signal can be removed without affecting the network activity.

In order for the network to learn, a teaching signal must be provided to the somatic compartment of the neurons in the visible population.
This signal nudges the somatic voltage of the visible neurons towards the appropriate activation at the appropriate times.
timing.
In turn, the soma targeting connections distribute the teaching signal spatially and temporally (given the delays) towards the latent layer.
Once the nudging is active, dendrite targeting connections learn to drive the dendritic voltage of postsynaptic partners towards the desired behavior.

%% file: sections/results.tex
\section{Results}\label{sec:results}

\subsection{Local plasticity enables complex sequence replay}

Following the developmental phase, the network can be trained to reproduce a complex spatio-temporal sequence.
Here, we train it to replay a long (compared to the neuronal time scales) non-Markovian pattern which serves as a mock-up of bird song.
For this we chose as a first example the beginning phrase of Beethoven's ``Für Elise'' (see \cref{fig:function}).
The visible population contains 13 neurons, one for each different pitch in the musical phrase.
The latent population contains 50 neurons for our baseline test.
To train the network in the replay of the melody, the visible population is repeatedly nudged towards a target activity pattern.

To observe the evolution of the network during learning, we interleave training with validation phases where the nudging to the visible population is removed for one pattern duration and the free network activity is observed (\cref{fig:function}a).
This allows to compare the network activities in the nudged state with those in the free state and can be used to judge how well the network has already learned to replay the pattern on its own.
As dendrite targeting synapses start out as weak, the activity in the latent layer is also too weak to sustain itself and the visible activity without the nudging input.
Learning then shapes and strengthens latent activity towards an increasingly structured dynamical attractor that allows the network to keep replaying the sequence even when the teaching signal is removed.
For evaluating the quality of the pattern replay we use two measures: the \gls{mse} between the target pattern and the activity in the visible population (\cref{fig:function}b) and the correlation coefficient between the two (\cref{fig:function}c).
For a more detailed description of these performance measures, see the Methods \cref{sec:meth-perf}.
After training, a replay phase starts, in which nudging by the teacher is removed completely and the network is allowed to keep evolving in time.
This evolution includes further weight plasticity, but in the absence of a teaching signal to the visible population.

Our results in \cref{fig:function}b and c show that latent populations as small as 30 neurons are sufficient to learn a very accurate and stable replay of the chosen visible pattern.
Importantly, as our model is capable of distributing useful teaching signals throughout the latent population, it can fulfill its task with a much smaller pool of neurons compared to other models of  birdsong learning or reservoir computing \citep{maes2020learning,jaeger2007optimization,sussillo2009generating,nicola2017supervised}.
Furthermore, the high fidelity of the produced pattern over 100 replay episodes shows that the network does not unlearn even though it remains plastic during replay.

\subsection{Learning of latent connectivity enables sustained pattern replay}

To isolate and highlight the importance of learning in the latent population, we perform an ablation study.
We compare our model containing plastic all-to-all dendrite targeting connections to a restricted version in which we only allow learning of the dendrite targeting weights that connect to the visible population.
Note that this ablated network is still more powerful than a standard reservoir, as it allows learning of recurrent connections in the visible population; additional ablation of learning in these recurrent weights would recover the reservoir computing paradigm.
Reservoirs and similar networks are known to be sensitive to the initizalization of their latent weights~\citep{tanaka2019recent, kawai2019smallworld, maass2002realtime, dale2021reservoir}.
Therefore, to ensure that we initialize our ablated networks in a way that does not prevent them from solving the task, we record the final weight distribution in the fully-trained (non-ablated) networks and initialize the static weights in the ablated networks with weights drawn randomly from that distribution.
This fulfills the principle of random initialization of a reservoir, while at the same time simplifying the work of the reservoir, as the initial distribution of weights is known to be able to solve the problem.

\cref{fig:reservoirComparison} shows that while the reservoir-mimicking ablated model seems to converge faster during training, the full model, capable of learning all connections,
is able to reach a much lower replay errors.
The observed increase in convergence speed is a direct consequence of having already initialized the latent weights in a useful regime for the specific task to be solved.
In contrast, the full model starts with vanishing latent weights which need to be learned in the first place.
However, training speed is obviously irrelevant if the network ultimately fails to perform the sought task.
As seen in \cref{fig:reservoirComparison}c, while the full model is capable of self-sustained
replay of the target pattern, the ablated network quickly diverges from the target pattern following the removal of the teacher.

\subsection{\elise networks can reliably learn diverse temporal patterns}

\begin{figure}[t!]
    \centering
    \includegraphics[width=0.5\textwidth]{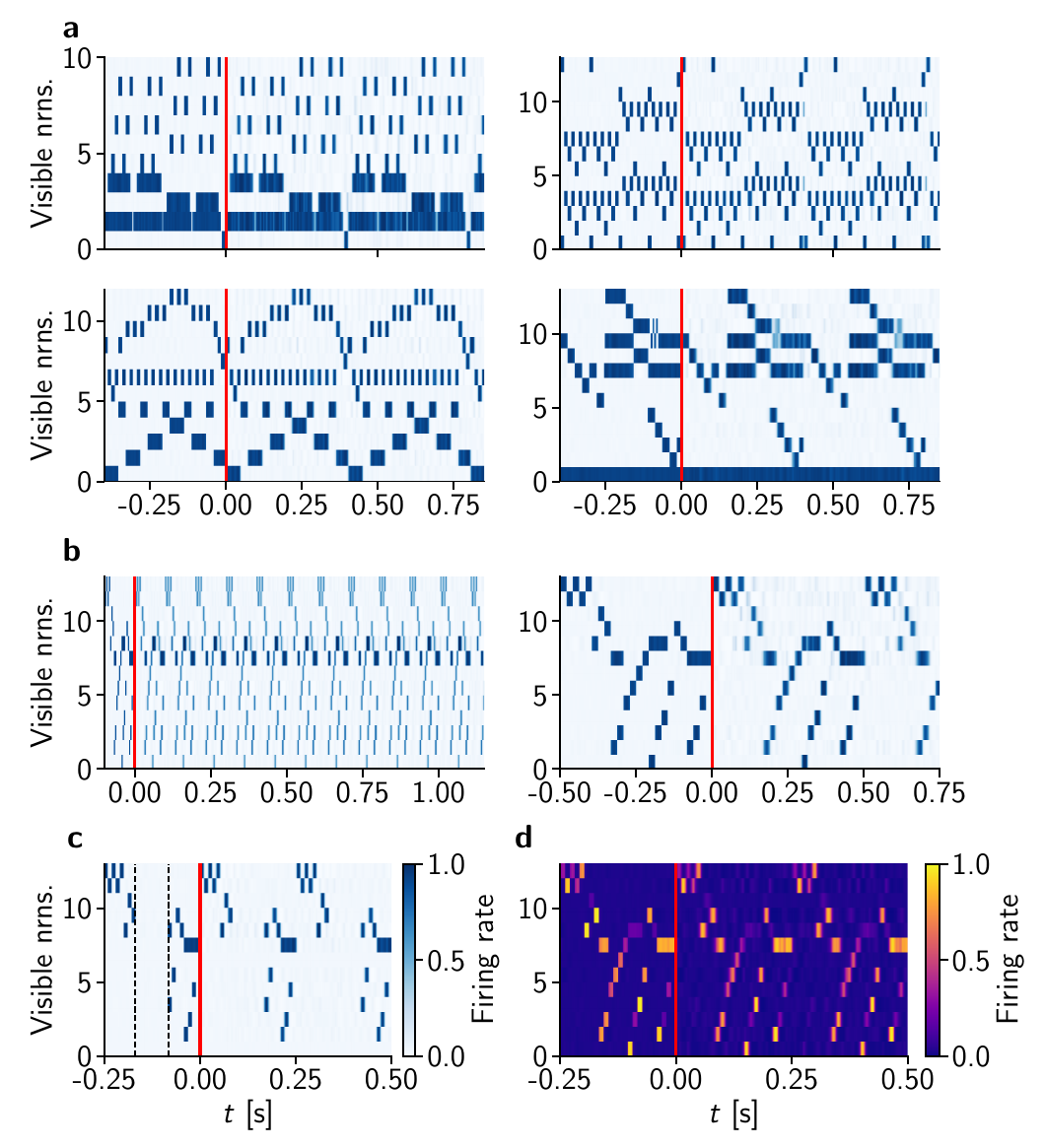}
    \caption{
      \tb{Networks are capable of learning a variety of patterns.}
      Individual plots show network activity after 2000 training cycles, with the red bar indicating the end of teacher nudging and the start of free replay. Network parameters are identical across different patterns.
      \tb{(a)} The first four patterns contained in our bespoke Bach preludes dataset.
      \tb{(b)} The ''Für Elise'' pattern at double (left) and half (right) speed, \tb{(c)} containing a large silent gap (between dashed lines) and \tb{(d)} composed of continuously valued (as opposed to binary) note intensities / targets.
    }
    \label{fig:showoff}
\end{figure}

So far we have shown that \elise networks are capable of learning a specific non-Markovian sequence, namely their namesake sonata. 
However, to be useful for general sequence learning, a system must be flexible with respect to sequence features such as tempo, length and pattern or sub-pattern structure, without the need for fine tuning. 
To test whether \elise learning generalizes to diverse complex sequences, we created a dataset based on the Well-Tempered Clavier preludes of Bach.  The dataset contains patterns composed of the first few measures of preludes 1-7. These pieces are non-Markovian, containing long repeating passages, and vary strongly in their timing and structure.
All networks contained 100 neurons in their latent population, were parameterized in the same way as in the previous tests, and trained as described in \cref{sec:meth-train}.
In (\cref{fig:showoff}a) we show that the \elise network learns to replay all these different sequences to a high degree of accuracy, thus showcasing the model's flexibility with respect to note lengths, note density and overall sequence structure.

A flexible sequence learning model should also be able to bridge different timescales, in the sense of adapting to short time scale sub-sequences as they are embedded within sequences of much longer time scales. Thus, while intuitively the size of the network determines an upper bound on overall sequence duration, below this bound the network should be able to learn sequences containing dynamics at very different time scales.
To probe this, we tested that identically parameterized and initialized networks are capable of learning the original "Für Elise" segment at half and double the original speed (\cref{fig:showoff}b).
Thanks to great diversity of delays and delay chains the latent population contains rich representation of past inputs at different temporal scales from which the visible layer can pick out relevant activities. 
This capability stands in contrast to clock-based approaches (such as \cite{maes2020learning}) in which pattern and sub-pattern durations must be fixed at the onset of learning, and the system clock would need to know the shortest relevant time scales from the get go.

A separate issue is whether \elise genuinely harnesses the long chains of delayed activities stored within the latent network to produce the target in the output.
It could be the case that the replay is driven mainly by visible to visible connections, with activity from the latent population acting as a mere gating mechanism.
To test this we trained a network on a pattern containing a long silence (\cref{fig:showoff}c). 
At $80$ ms, the period of silence is eight times longer than the neuronal time constants, and three times longer than the longest individual delay. 
Thus, the network needs to route activities via chains spanning multiple latent neurons to carry activities across the gap. 
Successful learning of this sequence supports the notion that the network makes extensive use of the activities stored in the latent population.

We also determined that \elise networks are capable of a further generalization. Thus far we have demonstrated the learning of patterns consisting of binary states, in which the teaching signal is composed of voltages corresponding either to very high or no firing.
But many sequences contain internal dynamics, meaning that elements within them are not only associated with a pitch and duration but also a continuous valued intensity.
Such an intensity dimension is very relevant for temporal pattern generation in the real world, such as for speech or (bird)song.
By training a network on a continuously valued pattern we show that \elise is capable of learning this intensity dynamics associated with a sequence, in addition to its spatial and temporal structure (\cref{fig:showoff}d).
This underscores the high fidelity of learning, as this sensitivity to intensity requires the weights to adapt to activities laying in the most reactive range of the (sigmoid) activation function.

\subsection{\elise networks can perform cued pattern recall and completion}

The ability to perform pattern completion -- i.e., full recall from partial information -- is an essential computational task that appears under different guises in different contexts, including sequence learning.
Pattern completion is highly relevant in real-world scenarios in which input (sensory data) is limited by noise and partial observability.
In the context of \elise, pattern completion manifests as the ability to recover a stored sequence from a partial nudging.

\begin{figure}[t!]
    \centering
    \includegraphics[width=0.5\textwidth]{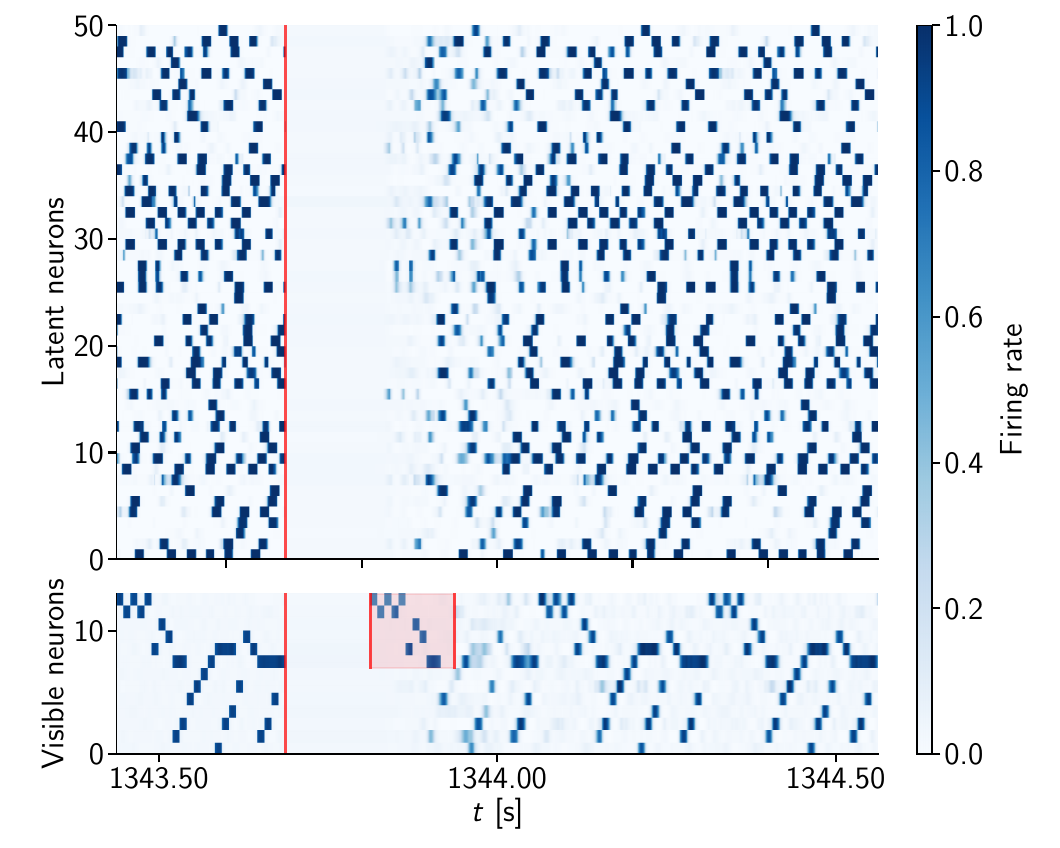}
    \caption{
        \tb{Networks perform cued pattern recall and completion.}
       At the end of training both the activity in the latent and visible population
       is reset.
       After a short silence period the network is nudged by cue comprised of a small subsection of the original pattern (shaded area).
       The cue is sufficient to reignite replay of the learned pattern in the network.
     }
    \label{fig:i_member}
\end{figure}

In \cref{fig:i_member} we show a test of the pattern completion capabilities for \elise networks. 
Following learning, the activities in both the visible and latent population are reset to zero, disrupting all dynamics. 
We then let the network stay in this silent state for half of a pattern duration (125 ms),
after which a subset of the visible layer is nudged with only a brief subsection of the full pattern. 
This brief and spatially limited cue is sufficient to reignite the network to replay the full-length pattern. 
Such a behavior speaks to the robustness of the dynamical attractor corresponding to the learned sequence.

\subsection{Simultaneous learning of multiple patterns}

\begin{figure*}[t]
    \includegraphics[width=1\textwidth]{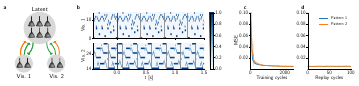}
    \caption{
        \tb{Simultaneous pattern learning in separate visible populations.}
        \tb{(a)} Two separate visible populations connecting to a single shared latent population. The visible populations do not directly innervate each other.
        \tb{(b)} Each visible population is given a teaching signal (Preludes 4 and 7 respectively). Their length ratio is irrational such that the overall pattern across both visible populations never repeats. Dashed lines indicate the individual pattern lengths.
        \tb{(c)} Performance on both patterns throughout training measured during validation intervals.
        \tb{(d)} Replay performance on both patterns following teacher removal.}
    \label{fig:multi}
\end{figure*}

Next, we investigate the multi-pattern learning capabilities of \elise networks. 
What we aim to show is how a single latent population can learn to simultaneously encode representations of very different patterns. 
The potential to learn multiple patterns within the same network greatly widens the range of applicability of \elise networks. 
In particular, it elevates the \elise model beyond the capabilities of models based on global clocks, as such models would struggle to find a common underlying basic rhythm among two independent sequences, especially in the case in which the timings of both sequences are incommensurate. 

We designed a scenario to test multi-pattern capabilities, in which two independent visible populations are coupled to a shared latent population (see \cref{fig:multi}a). 
The visible populations do not innervate each other, neither via somatic nor dendritic connections.
This ensures that the nudging information about both sequences arrives independently into a latent population, which in this case contains $200$ neurons.
Given this architecture, each visible layer is nudged with a different teaching signal, using two very distinct Bach prelude patterns.
Importantly, the patterns have different durations, with pattern lengths having a ratio equal to the golden ratio -- an irrational number. 
In this way, we make sure that there are no repetitions in the visible activity, and the network cannot merely learn a larger combined pattern. 
Despite this disjunction and their very different temporal structure, the latent population is able to support simultaneous learning and replay of both patterns (\cref{fig:multi}).
For this experiment to be successful the latent network must be storing the two patterns, without enabling pattern interference.

\subsection{Learning engenders robustness to internal variability and external disruption}

Aside from raw performance, we consider robustness to be an essential property of biologically plausible models.
The requirement of finely tuned parameters or disturbance-free operation is difficult to reconcile with the conditions under which biological neural networks must operate.

First, we demonstrate that our model is capable of accurate and stable sequence learning over a wide parameter range of $p$ and $q$, which control the developmental process and hence the connectivity of the nudging scaffold (\cref{fig:stability}a and b).
Next, we study the impact of a less dense dendrite targeting connectivity (\cref{fig:stability}c and d) by changing the proportion $r$ of non-zero and learnable connections 
(for more information about the sparsity see Methods \cref{sec:meth-sparse}).
While sparser networks require more training cycles to learn, networks with only \SI{40}{\percent} of all available connections are still able to learn and replay the pattern with low error (\cref{fig:stability}c).
Furthermore, we find that learning in even sparser networks is possible if we ensure that the dendrite targeting connections that follow the scaffold of nudging connections remain non-zero (\cref{fig:stability}d).
Since it is this soma targeting scaffold that propagates the teaching signals throughout the network, it appears plausible that the dendrite targeting connections that follow this scaffold are especially important for learning the task.

To model the effect of external disruptions to the visible population activity, we temporarily clamp the somatic voltages of the visible neurons to fixed values (\cref{fig:disruptedReplay}, see Methods \cref{sec:meth-disrupt} for details).
Although limited to the visible layer, this represents a strong disruption, as the visible layer plays a central role during learning, and clamping these neurons necessarily disturbs the behavior of the entire network.
For maximal impact, we clamp the voltage of the visible neurons either to their leak potential, modeling strong external suppression (\cref{fig:disruptedReplay}a and b), or to a high voltage above the leak potential, modeling strong external excitation (\cref{fig:disruptedReplay}c and d).
Plasticity is kept on during the clamping.
In both cases the network activity recovers (in all trials if only part of the pattern is disrupted and in most trials even when disrupting the entire pattern for one cycle) following a short transient phase during which the network settles back into its original behavior (see also Supplementary \cref{fig:disruptionSingleRuns}).
These experiments show that the dynamical attractors carved into the networks' phase space by the dendritic  learning are strong enough to withstand considerable disruptions.

%% file: sections/discussion.tex
\section{Discussion}
\FloatBarrier

In this work we describe \elise, an efficient and biologically plausible model of complex spatio-temporal sequence learning.
We take inspiration from biology in two ways.
First, we model structured neurons composed of a dendritic and a somatic compartment.
This allows synapses local access to multiple factors required for learning \cite{urbanczik2014learning}.
Second, we introduce structure to the network connectivity prior to learning, in a process mimicking early development.
Intuitively this propagates teaching signals (targets) throughout the network in a structured way, unlike in a randomly initialized network with dense connectivity.
This allows postsynaptic neurons to select those presynaptic partners that generate useful activity at the right moment in time.
We show that networks trained with \elise are able to learn and then replay a diverse range of complex non-Markovian sequences to a high accuracy while remaining parsimonious in their use of neuronal resources.
By comparing our model to a reservoir-like ablation, we show that plasticity in the latent connections is crucial for replay, as it carves out specific stable dynamical attractors in the space of latent neuron activities. Additionally, unlike reservoir models, \elise is not sensitive to initial conditions and does not require
fine-tuning of hyper-parameters~\citep{maass2002realtime, tanaka2019recent, kawai2019smallworld, dale2021reservoir}.
The model is capable of learning a wide range of complex sequences associated with different lengths, structure and dynamics.
Indeed a single network can simultaneously learn two such patterns presented at different input populations using only one shared latent population.
Further, we can show that the trained networks can perform cued pattern completion and recall as they are able to complete previously learned patterns from short snippets.
Finally, our model is robust both to noisy targets during training and disruptions to the visible activity during replay.

\subsection{Comparison to other approaches}
\elise compares favorably to other models of biologically plausible sequence learning.
Most notably, in comparison to others of its kind, our model is very resource efficient.
Requiring as little as 30 hidden neurons to learn a long non-Markovian sequence, it uses at least an order of magnitude fewer neurons than comparable biological sequence learning models~(\force \cite{sussillo2009generating}, \follow \cite{gilra2017predicting}, the ``clock chain'' approach in \cite{maes2020learning} and a spiking \htm \cite{bouhadjar2022sequence}).
As discussed above, we suggest that the main reason lies in the structured propagation of learning signals in \elise.
In our model this propagation is performed very efficiently via sparse, delayed soma targeting connections.
This use of delays is not plausible in models where learning relies either on instantaneous error signals (\force, \follow) or propagated errors (\rtrl \cite{zipser1989subgrouping}, \rflo \cite{murray2019local}).
Thus these models rely solely on the slow neuron dynamics and network size to create the required transients.
While the spiking \htm makes use of delays, this is only done to restrict learning between proximal neurons.

\begin{figure}
    \includegraphics[width=0.49\textwidth]{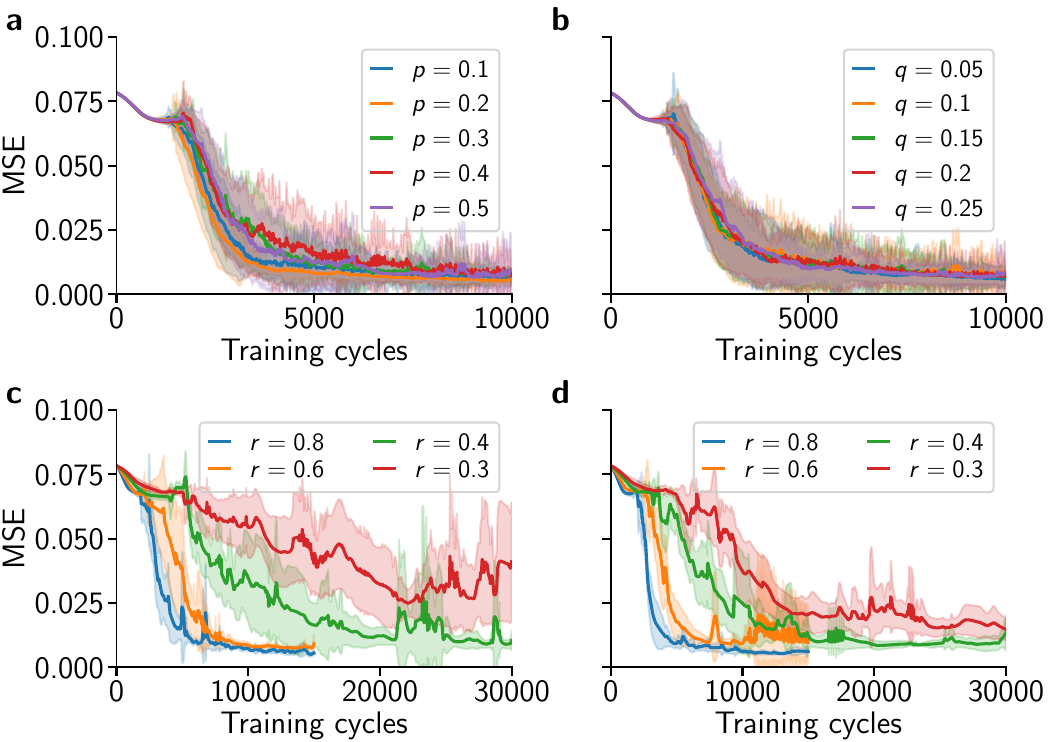}
    \caption{\tb{Robustness of learning with respect to changes in the network parameters.}
        \tb{(a)} Sweep over network parameter $p$. For each $p$ runs with a range of $q$ values and multiple seeds are averaged.
        \tb{(b)} Sweep over network parameter $q$. For each $q$ runs with a range of $p$ vlaues and multiple seeds are averaged.
        \tb{(c)} Training of networks with sparse dendrite targeting connectivity; $r$ represents the proportion of nonzero and plastic synapses out of all possible dendrite targeting connections in the network.
        \tb{(d)} Same as (c) but the dendrite targeting connections that match an already existing soma targeting connection are protected from sparsification.
    }
    \label{fig:stability}
\end{figure}

Further, our model is structurally simple.
The only distinguishing feature between the latent and visible populations is that the latter receives an external teaching input.
In contrast, other models of learning in recurrent networks (\force, \follow) make use of distinct functional units, such as for computing and transporting errors and signals, coupled with strong assumptions about the resulting connectivity.
The architectural simplicity of \elise is complemented by the simplicity and uniformity of its associated plasticity rule.
Learning is fully local and does not require the use of a global error signal (\force, \follow).
Finally, it does not require different phases (\follow) nor does it make biologically implausible assumptions about learning speed (\force).

\begin{figure}[t]
    \centering
    \includegraphics[width=0.49\textwidth]{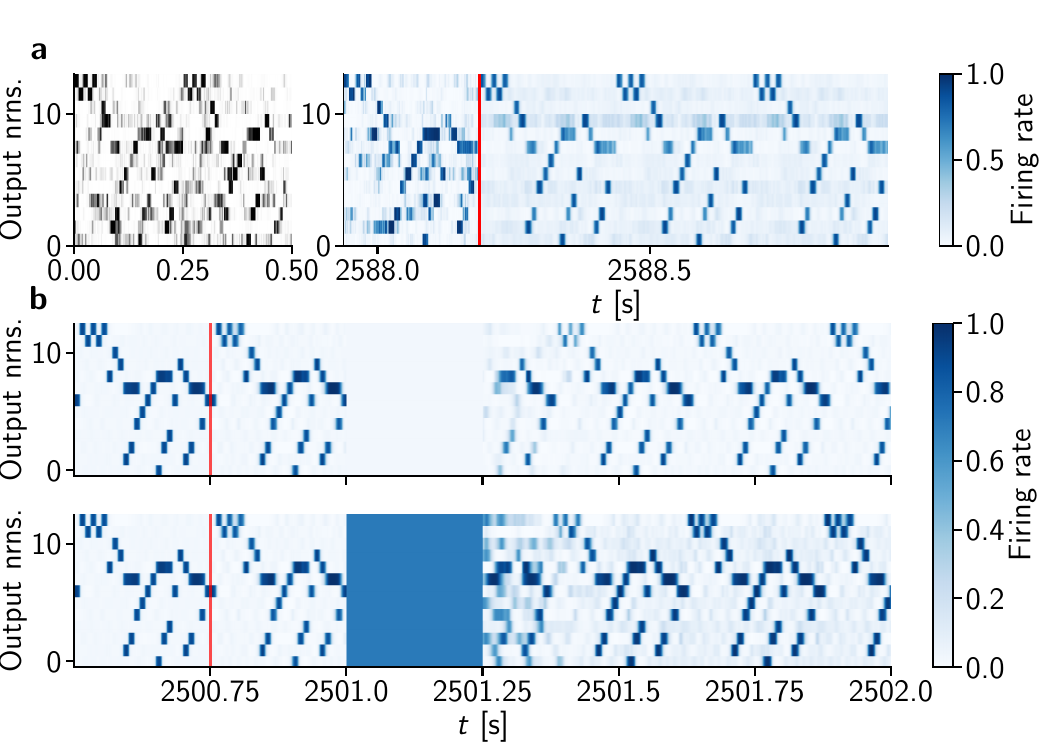}

    \caption{
        \tb{Networks are robust to disruptions during training and replay.}
        \tb{(a)} During training the teaching signal is corrupted by adding correlated noise (left).
        This noise is continuously sampled and thus different during every training cycle.
        Following teacher removal at the end of training, the network replays the denoised pattern.
        \tb{(b)} During replay the membrane potential of the visible population is clamped for one cycle such that it produces very low (top) or high (bottom) activity.
        Despite this, following the removal of the clamp the network recovers replay of the entrained pattern.}
    \label{fig:disruptedReplay}
\end{figure}

\subsection{Neural correlates}

While we do not model cortical areas explicitly, we can draw parallels between brain areas and their suggested functional roles and the components and dynamics of our model.
Here we compare our model to the neural mechanisms underlying birdsong learning in Zebra finches;
much like \elise, these birds learn a single song based on that of their teacher~\cite{bolhuis2006neural, hahnloser2010auditory}.
In our model, learning structures the latent activity until it becomes almost identical across replays of the sequence.
Qualitatively, this matches observations in zebra finches indicating that during learning the activity in pre-motor area \ra becomes increasingly self-similar~\cite{hahnloser2002ultrasparse,fee2004neural,lynch2016rhythmic}.
Furthermore, in our model the teaching signal is already structured at the onset of learning owing the soma targeting scaffold established during early development.
Similarly, evidence suggests that songbirds are born with a species-specific template that is refined though listening to the tutor~\cite{konishi1965role, gobes2007birdsong, bolhuis2015birdsong}.
Indeed, once the bird has learned the crystallized adult song, the removal of areas enconding the song (caudomedial nidopallium, caudomedial mesopalium) caused impairement of the tutor song recognition but not replay~\cite{gobes2007birdsong}.
Likewise, once our model has learned the sequence, the soma targeting connections no longer contribute to sequence replay and could be removed.

Finally, both our model and birdsong learning leverage sources of random unstructured activity to improve learning.
In our model the randomly initialized dendrite targeting connections initially propagate activity in a weak and unstructured way.
Learning allows \elise to single out and strengthen those connections that contribute to correct replay.
In birdsong learning variability is provided to pre-motor area \ra by the \lman~\cite{kao2005contributions, aronov2008specialized}.
It has been suggested that when variable \lman activity contributes to song replay, subcortical structures encoding reward signals induce plasticity in connections which then bias LMAN towards producing useful activity in the future~\cite{fee2011learning, gadagkar2016dopamine}.
These analogies between songbird learning and our model suggest that the interplay of structure and randomness is an essential feature of sequence learning systems.

\subsection{Open questions and future work}

Our model makes several simplifications, which we discuss in the following.
First, it does not explicitly obey Dale's law for the dendrite targeting synapses, simultaneously allowing both excitatory and inhibitory efferent connections.
However, the soma targeting interaction between pairs of pyramidal cells always follows a dual pathway, one direct and one mediated by an inhibitory interneuron.
While important for controlling the duration of this interaction \cite{wehr2003balanced, gabernet2005somatosensory},
the same mechanism can also be leveraged to implement a net inhibitory effect of dendrite targeting interactions,
as well as a seamless transition between net excitation and inhibition if plasticity is present in both pathways.
The implicitly assumed statistical balance between excitation and inhibition is supported by theoretical and experimental findings \cite{shadlen1994noise, vanvreeswijk1996chaos, shu2003turning, haider2006neocortical}

Second, we use rate-based neurons and forego the explicit modeling of spikes.
However, we expect \elise to transfer seamlessly to the spike-based regime.
The activity in our latent layer is strongly reminiscent of synfire braids and polychronization in spiking networks
\cite{abeles1982local, abeles1991corticonics, bienenstock1995model, izhikevich2006polychronization}
Moreover, our specific arrangement with paired feedforward inhibition exhibits the particularly appealing feature of convergence towards stable propagation \cite{diesmann1999stable}, with particular robustness towards variability in the neuronal substrate
\cite{kremkow2010gating, petrovici2014characterization}.

We believe this work inspires three further avenues of research.
First, the full computational performance of \elise is yet to be explored.
For example, here we are mainly focusing on learning a single non-Markovian sequence.
When we explored learning multiple sequences (\cref{fig:multi}) we did it in such a way that both sequences are reproduced simultaneously.
However, we expect our setup to also be applicable in scenarios where multiple patterns can be learned and reproduced by giving the appropriate cue to the network.
This would require larger networks and a training procedure that teaches the network all patterns at the same time, to prevent
catastrophic forgetting.
Second, it would be interesting to investigate more closely the link to the biological correlates of sequence learning.
One approach could be to explicitly model the functional properties of neuronal populations involved in sequence learning.
For example, one could investigate how the explicit inclusion of an external area providing random input (like \lman) or clock like input (like HVC) would affect learning in the model.
Alternatively, one could compare how our model networks and biological brains respond to the same kinds of disruptions such as cutting certain connections, injecting external currents or optogenetically activating individual populations.
Finally, we believe that the inherent resource efficiency, locality and robustness to disturbances make \elise an interesting candidate for deployment on neuromorphic systems and small-scale robotic platforms.
Especially in energy-constrained environments, for example on a small robotic system with a fixed motor activation sequence for movement, the small but also resilient network architectures learnt by \elise could provide tangible benefits compared to commonly deployed solutions.

%% file: sections/methods.tex
\section{Methods}

\subsection{Neuron model}\label{sec:meth-nrn}

We model our neurons as two compartments (a somatic and a dendritic one) which are both leaky integrators.
The compartments are coupled with a conductance $g_\den$ but only the dendrite influences the soma and not the other way around.
The dendritic membrane voltage $v$ is described as
\begin{align}
    C^\den_\text{m}\dot v &= - g_\leak \left(v - E_\leak\right) + I^\den_\syn
\end{align}
where $C^\den_\text{m}$ is the capacitance of the dendritic compartment, $g_\leak$ is its leak conductance and $E_\leak$ the resting potential.
The synaptic currents invoked by the incoming dendrite targeting connections are subsumed in $I^\den_\syn$.
The somatic membrane voltage $u$ is modeled as
\begin{align}
    C^\som_\text{m} \dot u &= - g_\leak \left(u - E_\leak\right) + g_\den \left(v - u\right) + I^\som_\syn
\end{align}
where $C^\som_\text{m}$ is the capacitance of the somatic compartment, $g_\leak$ is its leak conductance, $E_\leak$ the resting potential and $g_\den$ is the conductance coupling the dendrite to the soma.
The synaptic currents invoked by the incoming soma targeting connections are subsumed in $I^\som_\syn$.
The firing rate $r$ of the neuron is a logistic
function of the somatic voltage
\begin{align}
    r = \varphi\left(u\right) = \frac{1}{1 + \exp\left(a \left(b - u\right)\right)}
\end{align}
with parameters $a = 0.3$ and $b= -58$.
All additional parameters can be found in \cref{tab:sim_params} in the Supplement.

\subsection{Developmental setup}\label{sec:meth-dev}
In order to mimic biological developmental processes, we use a stochastic algorithm that relies only on local information.
We start out with two populations of neurons that have no synaptic connections, a latent and a visible population.
Each neuron in the visible population receives a somatic nudging connection from a teacher.
From this starting scenario we run a stochastic iterative algorithm to form the scaffold of soma targeting connections.

\begin{figure}[t]
    \centering
    \includegraphics[width=0.49\textwidth]{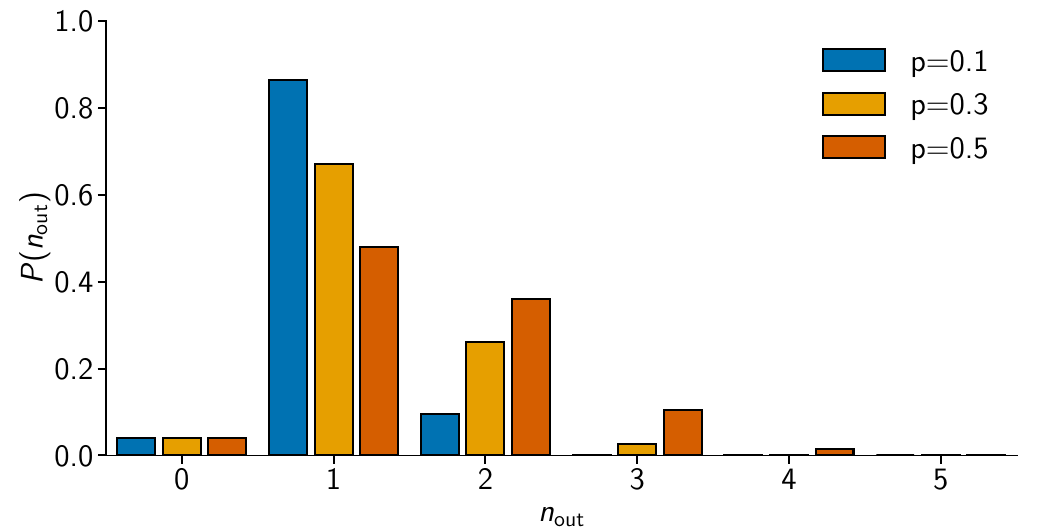}
    \caption{\tb{Distribution of $\bf{P(n_{out})}$.} Probability distribution of the number of outgoing connections $n_\mathrm{out}$ formed by a neuron depending on the parameter $p$. It can be described as a second order non-homogenous geometric distribution which is sparser that a first-order homogenous one.
    }
    \label{fig:p_distribution}
\end{figure}

The algorithm generates outgoing nudging (soma targeting) connections for those neurons which have at least one incoming nudging connection.
This ensures that only those neurons receiving teaching input themselves teach others.
Initially, the only neurons with incoming nudging connections are those in the visible layer.
During the wiring process, one neuron $i$ is randomly chosen from the set of neurons that receive nudging inputs and have not yet formed any outgoing connections.
This neuron $i$ makes X ($X\geq0$) outgoing nudging connections.
The probability to form zero connections is $p_0$, with a default of $p_0 = 0.04$.
For $X>0$, the first outgoing connection is made with probability $1 - p_0$.
Afterwards, the neuron develops additional outgoing connections with probability $p^{n_{\mathrm{out}}}$.
Here, $p = 0.2$ by default and $n_{\mathrm{out}}$ is the number of already existing outgoing connections.
Note that this is an iterative process, once the check if $n$ connections should be made is successful, a new random number is drawn and the check for $n + 1$ connections is performed.
Hence, the probability of $X=k$ is (see \Cref{fig:p_distribution}):

\begin{align}
    P(X=k) =
\begin{cases}
    p_0 & \quad \text{if } k = 0\\
    (1-p_0) p^{\frac{k^2 - k}{2}} (1-p^k)   & \quad \text{if } k \geq 1
\end{cases}
\end{align}

To find post-synaptic partners for the $X$ connections, for each of the $X$ connections a neuron $j$ from the latent population is selected randomly.
In the case that $i = j$ a new neuron is drawn as self-nudging is not possible.
Otherwise the probability for that neuron $j$ accepts the incoming connection is $q^M$ where $M$ is the number of previously accepted incoming connections (default: $q = 0.15$).
Therefore, a neuron that does not have incoming connections yet, will always accept its first connection.
If neuron $j$ rejects the connection, a new random neuron is selected.
Once a neuron $j$ accepts the connection, it is added to the set of neurons that must form outgoing connections, if it is not already in there.
When all $X$ connections are formed, neuron $i$ is permanently removed from the set of neurons that form outgoing connections, a new $i$ is chosen and the process repeats.
The setup algorithm terminates once there are no more neurons left in the set of neurons wanting to form outgoing connections.
Intuitively, we can think of $p$ controlling the density and $q$ the uniformity of the scaffold.

After the setup of the soma targeting nudging scaffold, the dendrite targeting connections are initialized as an all-to-all (including visible and latent population) weight matrix.
The individual weight values are drawn from a Gaussian
distribution with configurable means and standard deviations.
In particular, means and standard deviations can be set to different values for the 4 different subsets of connections: from visible to latent, from visible to visible, from latent to latent and from latent to visible.
By default, the mean is at zero and $\sigma = 0.5$.
For the experiments with sparse dendritic connections, the whole all-to-all weight matrix exists, but some of the connection weights are clamped to zero.

\subsection{Synaptic connections}\label{sec:meth-syn}
There are two types of synaptic connections: the conductance-based soma targeting (``nudging'') connections and the current-based dendrite targeting connections.

\subsubsection{Dendrite targeting connections}
The dendritic connection from neuron $i$ to neuron $j$ is characterized by its synaptic weight~$w^\den_{ji}$ and its transmission delay $d\,^\den_{ji}$.
All dendritic connections originating from neuron $i$ share the same $d\,^\den_i$ as a model for the delay introduced by the length of the axon.
The values of the dendritic delays are determined at network initialization and are multiples of the simulation time step ($dt = \SI{0.1}{ms}$).
They are drawn from a uniform distribution between $[d\,^\den_\text{min}, d\,^\den_\text{max}]$ with $d\,^\den_\text{min} = \SI{5}{ms}$ and $d\,^\den_\text{max} = \SI{15}{ms}$ in our simulations.
The summed synaptic current onto the dendritic compartment of neuron $j$ is then
\begin{align}
    I^\den_{\syn,j}\left(t\right) = \sum_i w^\den_{ji}\, r_i\left(t - d\,^\den_i\right)
\end{align}
with $r_i$ representing the presynaptic firing rate.

\subsubsection{Soma targeting connections}
The nudging scaffold consists of sparse, conductance-based soma targeting connections.
If the developmental algoritm in \cref{sec:meth-dev} sets up a nudging connection from neuron $i$ to neuron $j$, this nudging connection consists of an excitatory and an inhibitory connection.
The inhibitory connection is assumed to be produced via an inhibitory interneuron which is not explicitly modeled here (see \cref{fig:development}c).
Due to this ``additional length in signal transmission path'' the inhibitory connection is assumed to have a larger delay than its corresponding excitatory one.
This is modeled by the transmission delay $d\,^\som_{\inh,i}$ always being larger by a fixed amount $\delta$ $d\,^\som_{\inh,i} = d\,^\som_{\exc,i} + \delta$.
In our case $\delta = \SI{25}{ms}$.
The delay of all excitatory outgoing nudging connections formed by a neuron $i$ $d\,^\som_{\exc,i}$ is a multiple of the simulation time step $\Delta t$ and is drawn from a uniform distribution between $[d\,^\som_\text{min}, d\,^\som_\text{max}]$ with $d\,^\som_\text{min} = \SI{5}{ms}$ and $d\,^\som_\text{max} = \SI{15}{ms}$ in our simulations.
As all nudging connections have the same strength, the somatic weight matrix $w_{ji}^{som}$ is $1$ where a connection exists and $0$ otherwise.
The summed synaptic current onto the somatic compartment of neuron $j$ is
\begin{equation} \label{eq:i_syn_som}
    \begin{aligned}
      I^\som_{\syn,j} \left(t\right) =  \sum_i w_{ji}^{som}  [ &g_{\exc,i}(t) \left(E_\exc - u_{j}(t)\right) \\
      + &g_{\inh,i}(t) \left(E_\inh - u_{j}(t)\right)]
    \end{aligned}
\end{equation}

where $E_\exc$ and $E_\inh$ are the excitatory and inhibitory reversal potential.
The synaptic conductances $g_{\exc,i}(t)$ and $g_{\inh,i}(t)$ depend on the (delayed) firing rate $r_i$ of the pre-synaptic neuron $i$:
\begin{align}
    g_{\exc,i}\left(t\right) =
    \begin{cases}
        g_{\exc,0}\,\, r_i\left(t - d\,^\som_{\exc,i}\right) & \quad \text{if } r_i\left(t - d\,^\som_{\exc,i}\right) > \varphi(E_\leak)\\
        g_{\exc,0}\,\, \varphi\left(E_\leak\right) & \quad \text{otherwise}
    \end{cases}
\end{align}
\begin{align}
    g_{\inh,i}\left(t\right) =
    \begin{cases}
        g_{\inh,0}\,\, r_i\left(t - d\,^\som_{\inh,i}\right) & \quad \text{if } r_i\left(t - d\,^\som_{\inh,i}\right) > \varphi(E_\leak)\\
        G_{\inh,0}\,\, \varphi\left(E_\leak\right) & \quad \text{otherwise}
    \end{cases}
\end{align}
In the case where the nudging (i.e.\ the teacher) is removed for the visible neurons (typically during replay) the conductances $g_\exc$ and $g_\inh$ are set to zero.

\subsubsection{Teaching synapses}
The visible neurons receive their somatic inputs from an external teacher.
While the synaptic current $I^\som_\syn$ induced by these teaching synapses is of the same form as described in \cref{eq:i_syn_som}, the values of $g_\exc(t)$ and $g_\inh(t)$ are determined (without any delay) by the target pattern.

The target pattern consists of an array of values between 0 and 1 for each visible neuron, which indicates the activity over the pattern duration (a 0 indicating minimal firing rate and a 1 indicating maximal firing rate).
In most of our examples the pattern consists exclusively of the discrete values of 0 and 1 but in principle intermediate target activities are possible. Such an example is explicitly shown in \cref{fig:showoff}d.
The target activities are mapped to target voltages $u_\tgt(t)$ and we choose to map a target activity of 0 to $E_\leak$ and 1 to $E_\leak + 20$.

For a neuron model as described in \cref{sec:meth-nrn,sec:meth-syn} we know from \cite{urbanczik2014learning} that if the somatic potential is equal to a so-called ``matching potential''
\begin{align}
    u_\text{M} = \frac{g_\exc\,E_\exc + g_\inh\,E_\inh}{g_\exc + g_\inh},
\end{align}
then the net-effect of the conductance-based synapses to the soma amounts to zero and the somatic voltage $u$ tend towards the same value, even if the somatic connections are removed.
This is exactly the behavior we want from our teaching synapses:
Once the somatic voltage reaches its target, we want to be able to take the teaching synapses away and the neuron should still be able to produce the same (target) somatic voltage.
Therefore, the target voltage for the visible neurons $u_\tgt$ must correspond to the matching potential $u_\text{M}$ induced by the conductance based teaching synapses.
From this we can calculate the resulting synaptic conductances $g_\exc$ and $g_\inh$ that produce the desired matching potential by setting
\begin{align}
    u_\tgt = \frac{g_\exc\,E_\exc + g_\inh\,E_\inh}{g_\exc + g_\inh}
\end{align}
and solving for the conductances:
\begin{align}
    g_\exc &= \frac{g_\inh\left(E_\inh - u_\tgt\right)}{u_\tgt - E_\exc}\\
    g_\inh &= \frac{g_\exc\left(u_\tgt - E_\exc\right)}{E_\inh - u_\tgt}
\end{align}
As this is underdetermined, we impose another condition that enables us to balance the strength of the teacher nudging with the strength of leak and dendritic compartment by setting
\begin{align}
    g_\exc + g_\inh = \frac{\lambda}{1-\lambda} \left(g_\leak + g_\den\right)
\end{align}
with $\lambda$ as a scaling parameter that we typically set to $\lambda = 0.6$.
From this we can calculate the conductances of the teaching synapses dependent on the target voltage $u_\tgt$ given by the target pattern:
\begin{align}
    g_\exc\left(u_\tgt\right) &= \frac{\lambda}{1-\lambda} \left(g_\leak + g_\den\right) \frac{\left(E_\inh - u_\tgt\right)}{E_\inh - E_\exc}\\
    g_\inh\left(u_\tgt\right) &= \frac{\lambda}{1-\lambda} \left(g_\leak + g_\den\right) \frac{\left(u_\tgt - E_\exc\right)}{E_\inh - E_\exc}
\end{align}

\subsection{Plasticity}\label{sec:meth-plast}
Only the dendrite targeting synapses are plastic.
They learn with a local error-correcting learning rule similar to \cite{urbanczik2014learning}:
\begin{align}
    \dot w^\den_{ji} = \eta \, \left[\varphi\left(u_j\right) - \varphi\left(v^*_j\right)\right] \bar r_i
\end{align}
where $\eta$ is the learning rate and is different for the connections among the visible neurons ($\eta_o$) and all other connections ($\eta_l$).
$v^*$ is a scaled version of the dendritic voltage that corresponds to the voltage in the soma that a dendritic potential $v$ would elicit if the soma did not receive any additional input:
\begin{align}
    v^* = \frac{1}{g_\leak + g_\den} \, \left[g_\leak E_\leak + g_\den v\right]
\end{align}
$\bar r$ is a low-pass filter of the delayed pre-synpatic rate $r_i$ and represents the impact of the pre-synaptic rate on the dendritic potential
\begin{align}
    \dot{\bar{r}}_i\left(t\right) = - g_\leak \bar{r}_i\left(t\right) + \frac{g_\leak g_\den}{g_\leak + g_\den} r_i\left(t - d\,^\den_{i}\right)
.\end{align}

\subsection{Training setup}\label{sec:meth-train}
The network is trained by repeatedly nudging the visible neurons towards the target pattern.
Typically the target is presented for 10000 cycles, with the exception of the sparsity experiments which required longer training times.
There is no reset of voltages or rates between the cycles.
Interleaved into the training cycles, typically at intervals of 20 training cycles, there is one replay cycle to quantify training progress.
For these interleaved ``validation'' cycles the teacher nudging of the visible neurons is removed.
Weight updates continue during these validation cycles.

After the training cycles are completed, the replay phase, typically 100 cycles, starts.
There is no reset of voltages or rates at the switch between training and replay.
During the first three replay cycles the nudging is still active. Afterwards it is switched off and the network activity evolves without any outside influence.
Learning is always on during the replay phase, but since the nudging of the visible neurons is off, the weights do not change strongly.
We keep the learning switched on to demonstrate that the system, if left alone, does not unlearn.

\subsubsection{Cued pattern recall and completion}\label{sec:meth-recall}

In \cref{fig:i_member} we test the networks associative memory capabilities.
Following training completion we remove the teacher and reset all of the network activites including those ``stored'' in the delayed connections. Following a short silence we nudge the network with a cue.
The cue is a short segment of a part of the pattern.

\subsubsection{Disruptions}\label{sec:meth-disrupt}

To disrupt learning as show in \cref{fig:disruptedReplay} a Ornstein-Uhlenbeck noise was added to target voltages.
\begin{equation}
\dot{U} = \frac{- U(t)}{\tau} + \sigma \sqrt{\frac{2}{\tau}} \, \xi
\end{equation}

\begin{equation}
\xi = \mathcal{N}(0, 1) \quad \langle \xi(t) \xi(t') \rangle = \delta(t - t')
\end{equation}
Where $\tau$ is the noise time constant and $\sigma$ controls the strength of the noise.
The noise was continuously resampled throughout training and thus different for every presentation of the target.

For the experiments shown in \cref{fig:disruptedReplay}b the activity of the neurons in the visible population is disturbed during the replay phase.
This is achieved by clamping the somatic voltage of the disrupted neurons to a fixed value $u(t) = u_\text{disrupt}$ for all times $t$ within a given interval.
To model a disruption that silences the visible neurons $u_\text{disrupt} = E_\leak$ is chosen.
Conversely, to mimic a disruption that makes the visible neurons produce strong ``wrong'' activity we set $u_\text{disrupt} = E_\leak + 15$.

\subsubsection{Sparsity}\label{sec:meth-sparse}
The experiments in \cref{fig:stability}c and d investigate the impact of sparser dendrite targeting connectivities.
At the initialization of the dendrite targeting weight matrix, each entry is set to zero with the probability of $1 - r$, with $r$ being the chosen connection density of the dendrite targeting connections.
For \cref{fig:stability}d we protect those pairs of pre- and post-synaptic neurons from having their dendrite targeting weight nullified that also have a soma targeting connection.

\subsection{Performance measures}\label{sec:meth-perf}
There are several possible measures to compare the similarity of the temporal sequence produced by the visible population and the target sequence.
We use both the \gls{mse} and the correlation coefficient $\text{Coeff}$.
The correlation coefficient, specifically the Pearson correlation coefficient, is defined as the normalized covariance $\text{cov}$ between two signals $x$ and $y$:
\begin{align}
    \text{Coeff} = \frac{1}{\sqrt{\text{cov}(x, x)\text{cov}(y,y)}}\text{cov}(x, y)
\end{align}
In our case the signals $x = [x_0, x_1, \dots, x_T]$ is made out of the visible rate of a visible neuron over time and $y$ the target for this neuron.
The covariance is defined as
\begin{align}
    \text{cov}(x, y) = \frac{1}{T} \sum_{t=0}^{T}\left(x_t - E[x]\right)\left(y_t - E[y]\right)
\end{align}
where $E[x]$ is the expectation value of the variable $x$ and $E[y]$ the expectation value of $y$.
We then average the $\text{Coeff}$ for all visible neurons to get the final performance score.
While the correlation coefficient has the advantage of having easily interpretable performace values (a score of 1 being the optimum), a major disadvantage of it is that it is insensitive to a downscaling of the signal, e.g.\ $\text{Coeff}(x, y) = \text{Coeff}(\frac{1}{10}x, y)$.

We therefore complement it with the \gls{mse} as an additional performance measure.
The \gls{mse} for each neuron is calculated as mean over the quadratic differences between visible rate and target for every time step:
\begin{align}
    \text{MSE}(x, y) = \frac{1}{T} \sum_{t=0}^{T}\left(x_t - y_t\right)^2
\end{align}
We then average the \gls{mse} of all visible neurons.
The advantage of this measure is that it is sensitive to a scaling of one of the signals, therefore, e.g., a weakening of the visible during replay can be detected.
However, interpreting the performance values is slightly less intuitive, because while it is clear that lower is better, the absolute values of the \gls{mse} are not directly meaningful.

The combination of two different performance measures of which one is sensitive and one insensitive to a scaling of signals is advantageous because it provides more insights into the networks behavior and gives hints to reasons for performance losses.
If there is, for example, a dip in performance during replay that is visible in only the scaling-sensitive measure but not in the other one, this hints to a weakening of the signal because of loss of activity in the network.
If however the dip is visible in both measures, then the performance loss must be caused by something else, i.e.\ a real decorrelation of the signal and the target.

\subsubsection{Time shifts in the visible activity}\label{sec:meth-shift}
During replay the visible pattern shifts slightly to later times with every repetition.
We need to take this into account when evaluating replay performance.
To evaluate replay performance we evaluate the performance for each pattern repetition seperately, i.e. we calculate e.g. $\text{MSE}(x, y)$ with $x$ and $y$ containing the amount of time steps of one pattern duration $T$.
For every replay repetition $r$ we calculate the performance $p_r$ by evaluating all possible pattern shifts $s$ (not more than one pattern length is allowed as shift) and taking the best one
\begin{align}
    p_r &= \min_{s \in [0, T)}\left(\text{MSE}\left(x_{r, s}, y\right)\right) \;\;\; \text{with} \\
    x_{r, s} &= x[\,r\, T + s, (r + 1)\,T + s]
\end{align}
where $x$ is the full recording of the visible activity over all replay repetitions, $y$ is the target sequence, $r$ the replay index, $s$ the applied shift and $T$ the pattern duration.
Due to the shift, the quality of the last replay repetition cannot be evaluated.

%% file: sections/ack.tex
We thank all members of the NeuroTMA and the CompNeuro labs for many insightful discussions.
Additionally, we are grateful to Richard Hahnloser and his lab for valuable feedback on learning in songbirds.
We gratefully acknowledge funding from the European Union for the Human Brain Project under the grant agreement 945539, the Swiss National Science Foundation (SNSF) for the ``Prospective coding with pyramidal neurons''-grant (310030L\_156863), the the Sinergia Grant ``Neural Processing of Distinct Prediction Errors: Theory, Mechanisms \& Interventions'' (CRSII5\_180316)
and the Horizon Europe grant agreement 101147319 (EBRAINS 2.0)
Our work has greatly benefited from access to the Fenix Infrastructure resources, which are partially funded from the European Union's Horizon 2020 research and innovation programme through the ICEI project under the grant agreement No. 800858.
Last but not least, we owe a particular debt of gratitude for the ongoing support from the Manfred Stärk Foundation.

%% file: sections/authcont.tex
KS, MAP and WS designed the model.
KS wrote the initial C code for the software simulation which was adapted by LK and then extended and ported to Python by BvH and TG.
The experiments were jointly designed by LK, KS, BvH, FB, WS and MAP and conducted by BvH, LK and KS.
All authors contributed to the writing of the manuscript.

%% file: sections/supplement.tex
\section*{Supplementary Information}

\subsection*{Pattern replay after disruptions}\label{sec:supp-disruptions}

\begin{figure}[h]
    \centering
    \includegraphics[width=1.0\textwidth]{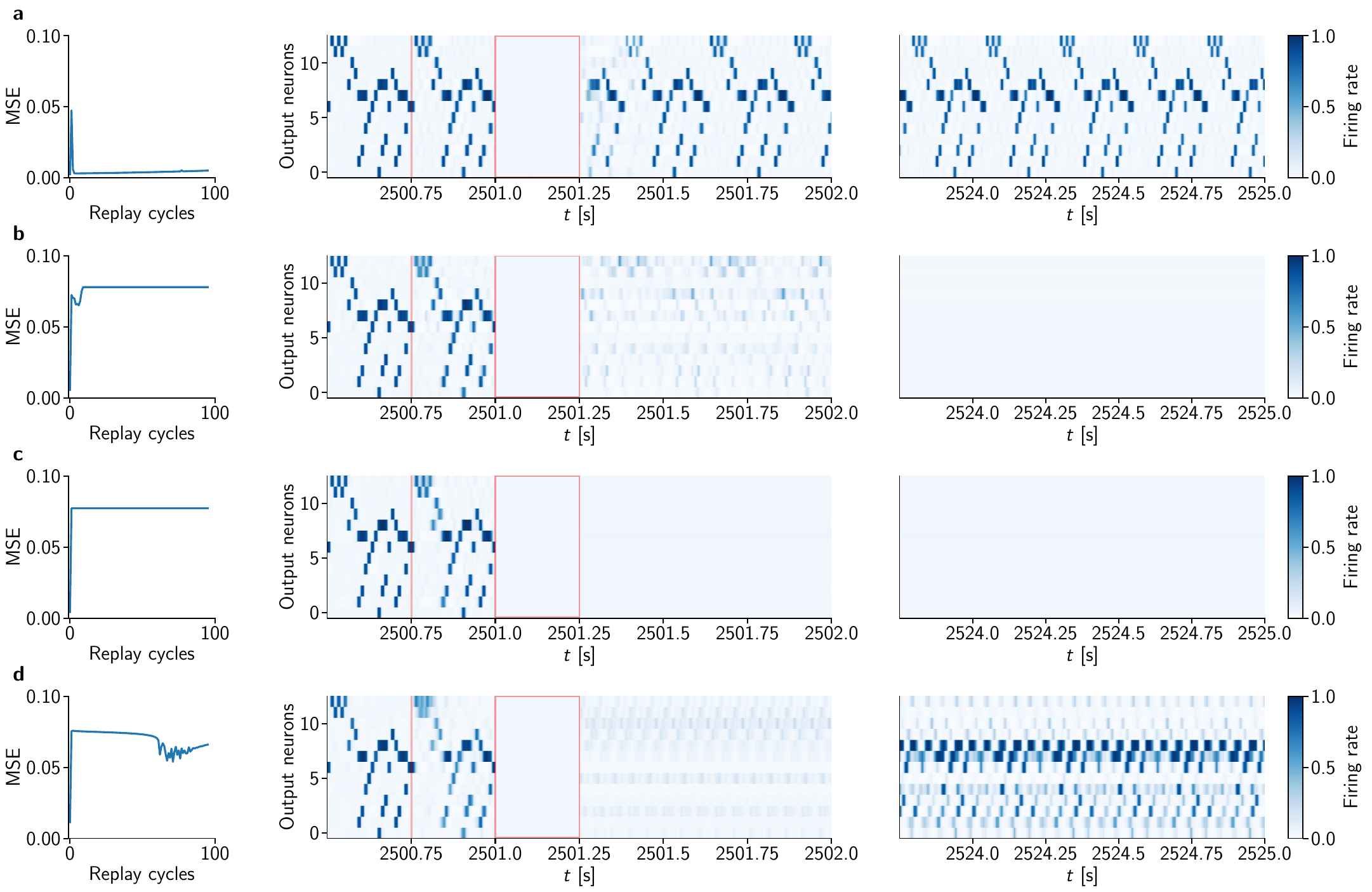}
    \caption{\textbf{Selected runs with disrupted replay of \cref{fig:disruptedReplay}.}
    Middle: Activity in output population at the beginning of the replay phase (red line marks the end of nudging).
    Right: Activity during the last replay cycles.}
    \label{fig:disruptionSingleRuns}
\end{figure}

\begin{figure}[h]
    \centering
    \includegraphics[width=0.9\textwidth]{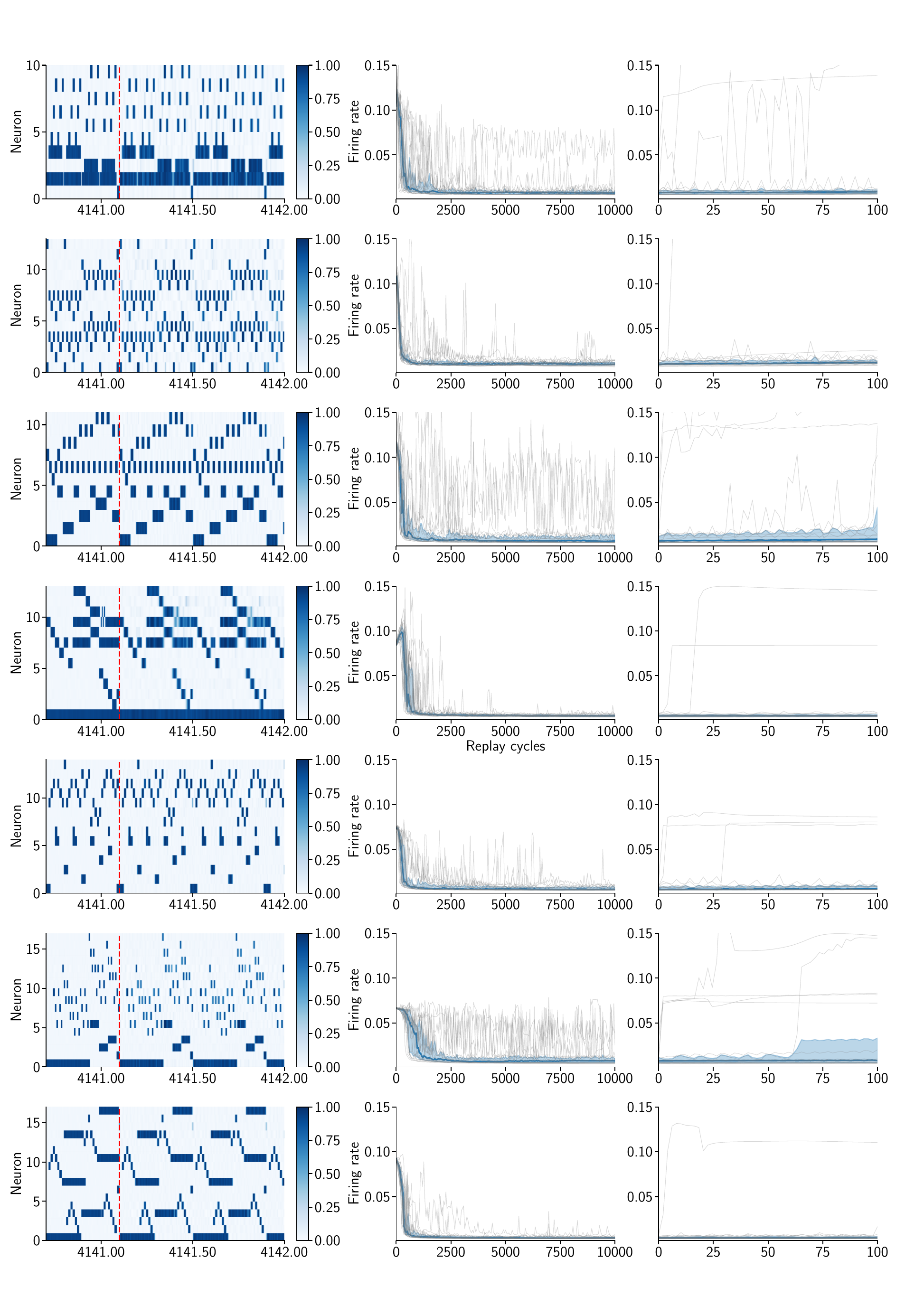}
    \caption{\textbf{Learning and replay of Bach Preludes}. Left: Activity in output population at the beginning of the replay phase (red line marks the end of nudging).
    Middle column: \gls{mse} error during training.
    Right column: \gls{mse} error during replay.}
    \label{fig:allPreludes}
\end{figure}

\newpage
\subsection*{Simulation parameters}\label{sec:supp-paramtables}
\begin{table}[h!]
    \caption{Parameters for simulations (part 1).}
    \centering
    \begin{threeparttable}
    \begin{tabular}{ccccc}
    \toprule
        & \cref{fig:function}  & \cref{fig:reservoirComparison} (c)\tnote{(1)} & \cref{fig:showoff} + \cref{fig:multi} & \cref{fig:i_member}  \\ \midrule
    $dt$~[ms]	& $10^{-1}$  & $10^{-1}$ & $10^{-1}$ & $10^{-1}$  \\
    Training cycles  & $10^{4} $  & $ 10^{4}$ & $5 \times 10^{3}$ & $5 \times 10^{3}$ \\
    Replay cycles  & $100 $  & $10$ & $100$ & $100$\\
    Validation interval  & $20$  & $20$ & $50$ & $50$\\ \midrule
    Latent neurons  & $[30, 50, 100]$  & $50$ & $100$ & $50$ \\
    $p$  & $0.2$  & $0.2$ & $0.2$ & $0.2$\\
    $p_0$  & $0.04$  & $0.04$ & $0.04$ & $0.04$\\
    $q$  & $0.15$  & $0.15$ & $0.15$ & $0.15$\\
    Dendritic sparsity  & all-to-all  & all-to-all & all-to-all & all-to-all \\ \midrule
    $C^\den_\text{m}$  & $1$  & $1$ & $1$ & $1$\\
    $C^\som_\text{m}$  & $1$  & $1$ & $1$ & $1$\\
    $E_\leak$  & $-70$  & $-70$ & $-70$ & $-70$\\
    $E_\exc$  & $0$  & $0$ & $0$ & $0$\\
    $E_\inh$  & $-75$  & $-75$ & $-75$ & $-75$\\
    $g_\leak$~[ms$^{-1}$]  & $0.1$  & $0.1$ & $0.1$ & $0.1$\\
    $g_\den$~[ms$^{-1}$]  & $2.0$  & $2.0$ & $2.0$ & $2.0$\\
    $g_{\exc,0}$~[ms$^{-1}$]  & $0.3$  & $0.3$ & $0.3$ & $0.3$\\
    $g_{\inh,0}$~[ms$^{-1}$]  & $6.0$  & $6.0$ & $6.0$ & $6.0$\\
    $a$\tnote{(2)}  & $0.3$  & $0.3$ & $0.3$ & $0.3$\\
    $b$\tnote{(2)}  & $-58$  & $-58$ & $-58$ & $-58$\\
    $d^\den$~[ms] & $\mathcal{U}[5, 15]$ & $\mathcal{U}[5, 15]$ & $\mathcal{U}[5, 25]$ & $\mathcal{U}[5, 25]$\\
    $d^\som_\exc$~[ms] & $\mathcal{U}[5, 15]$ & $\mathcal{U}[5, 15]$ & $\mathcal{U}[5, 25]$ & $\mathcal{U}[5, 25]$\\
    $\delta$~[ms] & 25 & 25 & 25 & 25\\ \midrule
    Init $\bm W_{\text{out}, \text{out}}$~[ms$^{-1}$] & $\mathcal{N}(0, 0.5)$ & $\mathcal{N}(0, 0.5)$ & $\mathcal{N}(0, 0.5)$ & $\mathcal{N}(0, 0.5)$\\
    Init $\bm W_{\text{out}, \text{lat}}$~[ms$^{-1}$] & $\mathcal{N}(0, 0.5)$ & $\mathcal{N}(0, 0.5)$ & $\mathcal{N}(0, 0.5)$ & $\mathcal{N}(0, 0.5)$\\
    Init $\bm W_{\text{lat}, \text{out}}$~[ms$^{-1}$] & $\mathcal{N}(0, 0.5)$ & $\mathcal{N}(0.19, 5.7)$ & $\mathcal{N}(0, 0.5)$ & $\mathcal{N}(0, 0.5)$\\
    Init $\bm W_{\text{lat}, \text{lat}}$~[ms$^{-1}$] & $\mathcal{N}(0, 0.5)$ & $\mathcal{N}(0.08, 6.1)$ & $\mathcal{N}(0, 0.5)$ & $\mathcal{N}(0, 0.5)$\\
    $\eta_\text{o}$~[ms$^{-2}$] & $10^{-4}$ & $4 \times 10^{-3}$ & $10^{-4}$ & $10^{-4}$\\
    $\eta_\text{l}$~[ms$^{-2}$] & $10^{-3}$ & $4 \times 10^{-2}$\tnote{(3)} & $10^{-3}$ & $10^{-3}$\\
    \bottomrule
    \end{tabular}
    \vspace{0.3cm}
    \begin{tablenotes}
    \footnotesize
    \item[1] Parameters for \cref{fig:reservoirComparison} (b) are the same as for \cref{fig:function}.
    \item[2] $r = \varphi(u) = \frac{1}{1 + \exp{a(b - u)}}$
    \item[3] In the reservoir-like network only the connections within the output population and the ones from the latent to the output population learn.
    \end{tablenotes}
    \end{threeparttable}
    \label{tab:sim_params}
\end{table}

\begin{table}[h!]
    \caption{Parameters for simulations (part 2).}
    \centering
    \begin{threeparttable}
    \begin{tabular}{ccccc}
    \toprule
        & \cref{fig:disruptedReplay} + \cref{fig:disruptionSingleRuns} & \cref{fig:stability} (a+b) & \cref{fig:stability} (c+d) & \cref{fig:allPreludes}\\ \midrule
    $dt$~[ms]	& $10^{-1}$  & $10^{-1}$ & $10^{-1}$ & $10^{-1}$ \\
    Training cycles  & $ 10^{4} $ & $ 10^{4} $ & up to $ 3 \times 10^{4} $ & $10^{4}$ \\
    Replay cycles  & $100$ & $100$ & $100$ & $100$ \\
    Validation interval  & $20$  & $20$ & $20$ & $50$ \\ \midrule
    Latent neurons  & $50$ & $50$ & $50$ & $100$ \\
    $p$  & $0.2$ & $0.1 - 0.5$ & $0.2$ & $0.2$ \\
    $p_0$  & $0.04$  & $0.04$ & $0.04$ & $0.04$ \\
    $q$  & $0.15$ & $0.05 - 0.25$ & $0.15$ & $0.15$ \\
    Dendritic sparsity  & all-to-all & all-to-all & $0.3 - 0.8$ & all-to-all \\ \midrule
    $C^\den_\text{m}$  & $1$  & $1$ & $1$ & $1$ \\
    $C^\som_\text{m}$  & $1$  & $1$ & $1$ & $1$ \\
    $E_\leak$  & $-70$  & $-70$ & $-70$ & $-70$ \\
    $E_\exc$  & $0$  & $0$ & $0$ & $0$ \\
    $E_\inh$  & $-75$  & $-75$ & $-75$ & $-75$ \\
    $g_\leak$~[ms$^{-1}$]  & $0.1$  & $0.1$ & $0.1$ & $0.1$ \\
    $g_\den$~[ms$^{-1}$]  & $2.0$  & $2.0$ & $2.0$ & $2.0$ \\
    $g_{\exc,0}$~[ms$^{-1}$]  & $0.3$  & $0.3$ & $0.3$ & $0.3$ \\
    $g_{\inh,0}$~[ms$^{-1}$]  & $6.0$  & $6.0$ & $6.0$ & $6.0$ \\
    $a$\tnote{(2)}  & $0.3$  & $0.3$ & $0.3$ & $0.3$ \\
    $b$\tnote{(2)}  & $-58$  & $-58$ & $-58$ & $-58$ \\
    $d^\den$~[ms] & $\mathcal{U}[5, 15]$ & $\mathcal{U}[5, 15]$ & $\mathcal{U}[5, 15]$ & $\mathcal{U}[5, 25]$ \\
    $d^\som_\exc$~[ms] & $\mathcal{U}[5, 15]$ & $\mathcal{U}[5, 15]$ & $\mathcal{U}[5, 15]$ & $\mathcal{U}[5, 25]$ \\
    $\delta$~[ms] & 25 & 25 & 25 & 25 \\ \midrule
    Init $\bm W_{\text{out}, \text{out}}$~[ms$^{-1}$] & $\mathcal{N}(0, 0.5)$ & $\mathcal{N}(0, 0.5)$ & $\mathcal{N}(0, 0.5)$& $\mathcal{N}(0, 0.5)$\\
    Init $\bm W_{\text{out}, \text{lat}}$~[ms$^{-1}$] & $\mathcal{N}(0, 0.5)$ & $\mathcal{N}(0, 0.5)$ & $\mathcal{N}(0, 0.5)$& $\mathcal{N}(0, 0.5)$\\
    Init $\bm W_{\text{lat}, \text{out}}$~[ms$^{-1}$] & $\mathcal{N}(0, 0.5)$ & $\mathcal{N}(0, 0.5)$ & $\mathcal{N}(0, 0.5)$& $\mathcal{N}(0, 0.5)$\\
    Init $\bm W_{\text{lat}, \text{lat}}$~[ms$^{-1}$] & $\mathcal{N}(0, 0.5)$ & $\mathcal{N}(0, 0.5)$ & $\mathcal{N}(0, 0.5)$& $\mathcal{N}(0, 0.5)$\\
    $\eta_\text{o}$~[ms$^{-2}$] & $10^{-4}$ & $10^{-4}$ & $10^{-4}$ & $10^{-4}$ \\
    $\eta_\text{l}$~[ms$^{-2}$] & $10^{-3}$ & $10^{-3}$ & $10^{-3}$ & $10^{-3}$ \\
    \bottomrule
    \end{tabular}
    \vspace{0.3cm}
    \begin{tablenotes}
    \footnotesize
    \item[2] $r = \varphi(u) = \frac{1}{1 + \exp{a(b - u)}}$
    \end{tablenotes}
    \end{threeparttable}
\end{table}

%% file: final.bib
@book{abeles1982local,
  title         = {Local Cortical Circuits: An Electrophysiological Study},
  author        = {Abeles, Moshe},
  year          = 1982,
  journal       = {Studies of Brain Function},
  publisher     = {Springer Berlin Heidelberg},
  doi           = {10.1007/978-3-642-81708-3},
  isbn          = 9783642817083,
  issn          = {0172-5742},
  url           = {http://dx.doi.org/10.1007/978-3-642-81708-3}
}

@book{abeles1991corticonics,
  title         = {Corticonics: Neural Circuits of the Cerebral Cortex},
  author        = {Abeles, Moshe},
  year          = 1991,
  month         = feb,
  publisher     = {Cambridge University Press},
  doi           = {10.1017/cbo9780511574566},
  isbn          = 9780511574566,
  url           = {http://dx.doi.org/10.1017/CBO9780511574566}
}

@article{adler2019somatostatinexpressing,
  title         = {Somatostatin-expressing interneurons enable and maintain learning-dependent sequential activation of pyramidal neurons},
  author        = {Adler, Avital and Zhao, Ruohe and Shin, Myung Eun and Yasuda, Ryohei and Gan, Wen-Biao},
  year          = 2019,
  month         = {apr},
  journal       = {Neuron},
  publisher     = {Elsevier},
  volume        = 102,
  number        = 1,
  pages         = {202--216},
  doi           = {10.1016/j.neuron.2019.01.036},
  issn          = {0896-6273},
  url           = {http://dx.doi.org/10.1016/j.neuron.2019.01.036}
}

@article{aronov2008specialized,
  title         = {A Specialized Forebrain Circuit for Vocal Babbling in the Juvenile Songbird},
  author        = {Aronov, Dmitriy and Andalman, Aaron S. and Fee, Michale S.},
  year          = 2008,
  month         = may,
  journal       = {Science},
  publisher     = {American Association for the Advancement of Science (AAAS)},
  volume        = 320,
  number        = 5876,
  pages         = {630--634},
  doi           = {10.1126/science.1155140},
  issn          = {1095-9203},
  url           = {http://dx.doi.org/10.1126/science.1155140}
}

@article{asabuki2022neural,
  title         = {Neural circuit mechanisms of hierarchical sequence learning tested on large-scale recording data},
  author        = {Asabuki, Toshitake and Kokate, Prajakta and Fukai, Tomoki},
  year          = 2022,
  month         = {jun},
  journal       = {PLOS Computational Biology},
  publisher     = {Public Library of Science San Francisco, CA USA},
  volume        = 18,
  number        = 6,
  pages         = {e1010214},
  doi           = {10.1371/journal.pcbi.1010214},
  issn          = {1553-7358},
  url           = {http://dx.doi.org/10.1371/journal.pcbi.1010214}
}

@article{bakkum2013tracking,
  title         = {Tracking axonal action potential propagation on a high-density microelectrode array across hundreds of sites},
  author        = {Bakkum, Douglas J and Frey, Urs and Radivojevic, Milos and Russell, Thomas L and M{\"u}ller, Jan and Fiscella, Michele and Takahashi, Hirokazu and Hierlemann, Andreas},
  year          = 2013,
  month         = {jul},
  journal       = {Nature Communications},
  publisher     = {Nature Publishing Group UK London},
  volume        = 4,
  number        = 1,
  pages         = 2181,
  doi           = {10.1038/ncomms3181},
  issn          = {2041-1723},
  url           = {http://dx.doi.org/10.1038/ncomms3181}
}

@article{bienenstock1995model,
  title         = {A model of neocortex},
  author        = {Bienenstock, Elie},
  year          = 1995,
  month         = {may},
  journal       = {Network: Computation in Neural Systems},
  publisher     = {Taylor \& Francis},
  volume        = 6,
  number        = 2,
  pages         = {179--224},
  doi           = {10.1088/0954-898x/6/2/004},
  issn          = {0954-898X},
  url           = {http://dx.doi.org/10.1088/0954-898x/6/2/004}
}

@article{bolhuis2006neural,
  title         = {Neural mechanisms of birdsong memory},
  author        = {Bolhuis, Johan J. and Gahr, Manfred},
  year          = 2006,
  month         = may,
  journal       = {Nature Reviews Neuroscience},
  publisher     = {Springer Science and Business Media LLC},
  volume        = 7,
  number        = 5,
  pages         = {347--357},
  doi           = {10.1038/nrn1904},
  issn          = {1471-0048},
  url           = {http://dx.doi.org/10.1038/nrn1904}
}

@article{bolhuis2015birdsong,
  title         = {Birdsong memory and the brain: In search of the template},
  author        = {Bolhuis, Johan J. and Moorman, Sanne},
  year          = 2015,
  month         = mar,
  journal       = {Neuroscience \& Biobehavioral Reviews},
  publisher     = {Elsevier BV},
  volume        = 50,
  pages         = {41--55},
  doi           = {10.1016/j.neubiorev.2014.11.019},
  issn          = {0149-7634},
  url           = {http://dx.doi.org/10.1016/j.neubiorev.2014.11.019}
}

@article{bouhadjar2022sequence,
  title         = {Sequence Learning, Prediction, and Replay in Networks of Spiking Neurons},
  author        = {Bouhadjar, Younes and Wouters, Dirk J. and Diesmann, Markus and Tetzlaff, Tom},
  year          = 2022,
  month         = jun,
  journal       = {PLOS Computational Biology},
  volume        = 18,
  number        = 6,
  pages         = {e1010233},
  doi           = {10.1371/journal.pcbi.1010233},
  issn          = {1553-7358},
  urldate       = {2022-07-25},
  editor        = {Kiebel, Stefan},
  langid        = {english}
}

@article{chang2021development,
  title         = {Development of auditory cortex circuits},
  author        = {Chang, Minzi and Kanold, Patrick O},
  year          = 2021,
  month         = {apr},
  journal       = {Journal of the Association for Research in Otolaryngology},
  publisher     = {Springer},
  volume        = 22,
  number        = 3,
  pages         = {237--259},
  doi           = {10.1007/s10162-021-00794-3},
  issn          = {1525-3961},
  url           = {http://dx.doi.org/10.1007/s10162-021-00794-3}
}

@article{chklovskii2004cortical,
  title         = {Cortical rewiring and information storage},
  author        = {Chklovskii, Dmitri B and Mel, BW and Svoboda, K},
  year          = 2004,
  month         = {oct},
  journal       = {Nature},
  publisher     = {Nature Publishing Group UK London},
  volume        = 431,
  number        = 7010,
  pages         = {782--788},
  doi           = {10.1038/nature03012},
  issn          = {0028-0836},
  url           = {http://dx.doi.org/10.1038/nature03012}
}

@article{cossart2022development,
  title         = {How development sculpts hippocampal circuits and function},
  author        = {Cossart, Rosa and Khazipov, Roustem},
  year          = 2022,
  month         = {jan},
  journal       = {Physiological Reviews},
  publisher     = {American Physiological Society Rockville, MD},
  volume        = 102,
  number        = 1,
  pages         = {343--378},
  doi           = {10.1152/physrev.00044.2020},
  issn          = {0031-9333},
  url           = {http://dx.doi.org/10.1152/physrev.00044.2020}
}

@article{dale2021reservoir,
  title         = {Reservoir computing quality: connectivity and topology},
  author        = {Dale, Matthew and O'Keefe, Simon and Sebald, Angelika and Stepney, Susan and Trefzer, Martin A},
  year          = 2021,
  month         = {dec},
  journal       = {Natural Computing},
  publisher     = {Springer},
  volume        = 20,
  pages         = {205--216},
  doi           = {10.1007/s11047-020-09823-1},
  issn          = {1567-7818},
  url           = {http://dx.doi.org/10.1007/s11047-020-09823-1}
}

@article{diesmann1999stable,
  title         = {Stable propagation of synchronous spiking in cortical neural networks},
  author        = {Diesmann, Markus and Gewaltig, Marc-Oliver and Aertsen, Ad},
  year          = 1999,
  month         = dec,
  journal       = {Nature},
  publisher     = {Springer Science and Business Media LLC},
  volume        = 402,
  number        = 6761,
  pages         = {529--533},
  doi           = {10.1038/990101},
  issn          = {1476-4687},
  url           = {http://dx.doi.org/10.1038/990101}
}

@article{fee2004neural,
  title         = {Neural mechanisms of vocal sequence generation in the songbird},
  author        = {Fee, Michale S and Kozhevnikov, Alexay A and Hahnloser, Richard HR},
  year          = 2004,
  month         = {jun},
  journal       = {Annals of the New York Academy of Sciences},
  publisher     = {Wiley Online Library},
  volume        = 1016,
  number        = 1,
  pages         = {153--170},
  doi           = {10.1196/annals.1298.022},
  issn          = {0077-8923},
  url           = {http://dx.doi.org/10.1196/annals.1298.022}
}

@article{fee2011learning,
  title         = {A hypothesis for basal ganglia-dependent reinforcement learning in the songbird},
  author        = {Fee, M. S. and Goldberg, J. H.},
  year          = 2011,
  month         = dec,
  journal       = {Neuroscience},
  publisher     = {Elsevier BV},
  volume        = 198,
  pages         = {152--170},
  doi           = {10.1016/j.neuroscience.2011.09.069},
  issn          = {0306-4522},
  url           = {http://dx.doi.org/10.1016/j.neuroscience.2011.09.069}
}

@article{gabernet2005somatosensory,
  title         = {Somatosensory integration controlled by dynamic thalamocortical feed-forward inhibition},
  author        = {Gabernet, Laetitia and Jadhav, Shantanu P and Feldman, Daniel E and Carandini, Matteo and Scanziani, Massimo},
  year          = 2005,
  month         = {oct},
  journal       = {Neuron},
  publisher     = {Elsevier},
  volume        = 48,
  number        = 2,
  pages         = {315--327},
  doi           = {10.1016/j.neuron.2005.09.022},
  issn          = {0896-6273},
  url           = {http://dx.doi.org/10.1016/j.neuron.2005.09.022}
}

@article{gadagkar2016dopamine,
  title         = {Dopamine neurons encode performance error in singing birds},
  author        = {Gadagkar, Vikram and Puzerey, Pavel A. and Chen, Ruidong and Baird-Daniel, Eliza and Farhang, Alexander R. and Goldberg, Jesse H.},
  year          = 2016,
  month         = dec,
  journal       = {Science},
  publisher     = {American Association for the Advancement of Science (AAAS)},
  volume        = 354,
  number        = 6317,
  pages         = {1278--1282},
  doi           = {10.1126/science.aah6837},
  issn          = {1095-9203},
  url           = {http://dx.doi.org/10.1126/science.aah6837}
}

@article{gilra2017predicting,
  title         = {Predicting Non-Linear Dynamics by Stable Local Learning in a Recurrent Spiking Neural Network},
  author        = {Gilra, Aditya and Gerstner, Wulfram},
  year          = 2017,
  month         = nov,
  journal       = {eLife},
  publisher     = {eLife Sciences Publications, Ltd},
  volume        = 6,
  pages         = {e28295},
  doi           = {10.7554/eLife.28295},
  issn          = {2050-084X},
  editor        = {Latham, Peter}
}

@article{gobes2007birdsong,
  title         = {Birdsong Memory: A Neural Dissociation between Song Recognition and Production},
  author        = {Gobes, Sharon M. H. and Bolhuis, Johan J.},
  year          = 2007,
  month         = may,
  journal       = {Current Biology},
  publisher     = {Elsevier BV},
  volume        = 17,
  number        = 9,
  pages         = {789--793},
  doi           = {10.1016/j.cub.2007.03.059},
  issn          = {0960-9822},
  url           = {http://dx.doi.org/10.1016/j.cub.2007.03.059}
}

@article{grosmark2016diversity,
  title         = {Diversity in neural firing dynamics supports both rigid and learned hippocampal sequences},
  author        = {Grosmark, Andres D and Buzs{\'a}ki, Gy{\"o}rgy},
  year          = 2016,
  month         = {mar},
  journal       = {Science},
  publisher     = {American Association for the Advancement of Science},
  volume        = 351,
  number        = 6280,
  pages         = {1440--1443},
  doi           = {10.1126/science.aad1935},
  issn          = {0036-8075},
  url           = {http://dx.doi.org/10.1126/science.aad1935}
}

@article{hahnloser2002ultrasparse,
  title         = {An ultra-sparse code underlies the generation of neural sequences in a songbird},
  author        = {Hahnloser, Richard HR and Kozhevnikov, Alexay A and Fee, Michale S},
  year          = 2002,
  month         = {sep},
  journal       = {Nature},
  publisher     = {Nature Publishing Group UK London},
  volume        = 419,
  number        = 6902,
  pages         = {65--70},
  doi           = {10.1038/nature00974},
  issn          = {0028-0836},
  url           = {http://dx.doi.org/10.1038/nature00974}
}

@article{hahnloser2010auditory,
  title         = {Auditory representations and memory in birdsong learning},
  author        = {Hahnloser, Richard H. R. and Kotowicz, Andreas},
  year          = 2010,
  month         = jun,
  journal       = {Current Opinion in Neurobiology},
  publisher     = {Elsevier BV},
  volume        = 20,
  number        = 3,
  pages         = {332--339},
  doi           = {10.1016/j.conb.2010.02.011},
  issn          = {0959-4388},
  url           = {http://dx.doi.org/10.1016/j.conb.2010.02.011}
}

@article{haider2006neocortical,
  title         = {Neocortical network activity in vivo is generated through a dynamic balance of excitation and inhibition},
  author        = {Haider, Bilal and Duque, Alvaro and Hasenstaub, Andrea R and McCormick, David A},
  year          = 2006,
  month         = {apr},
  journal       = {Journal of Neuroscience},
  publisher     = {Society of Neuroscience},
  volume        = 26,
  number        = 17,
  pages         = {4535--4545},
  doi           = {10.1523/jneurosci.5297-05.2006},
  issn          = {0270-6474},
  url           = {http://dx.doi.org/10.1523/jneurosci.5297-05.2006}
}

@article{harvey2012choicespecific,
  title         = {Choice-specific sequences in parietal cortex during a virtual-navigation decision task},
  author        = {Harvey, Christopher D and Coen, Philip and Tank, David W},
  year          = 2012,
  month         = {mar},
  journal       = {Nature},
  publisher     = {Nature Publishing Group UK London},
  volume        = 484,
  number        = 7392,
  pages         = {62--68},
  doi           = {10.1038/nature10918},
  issn          = {0028-0836},
  url           = {http://dx.doi.org/10.1038/nature10918}
}

@article{izhikevich2006polychronization,
  title         = {Polychronization: Computation with Spikes},
  author        = {Izhikevich, Eugene M.},
  year          = 2006,
  month         = feb,
  journal       = {Neural Computation},
  publisher     = {MIT Press - Journals},
  volume        = 18,
  number        = 2,
  pages         = {245--282},
  doi           = {10.1162/089976606775093882},
  issn          = {1530-888X},
  url           = {http://dx.doi.org/10.1162/089976606775093882}
}

@article{jaeger2007optimization,
  title         = {Optimization and Applications of Echo State Networks with Leaky- Integrator Neurons},
  author        = {Jaeger, Herbert and Luko{\v s}evi{\v c}ius, Mantas and Popovici, Dan and Siewert, Udo},
  year          = 2007,
  month         = apr,
  journal       = {Neural Networks},
  series        = {Echo {{State Networks}} and {{Liquid State Machines}}},
  volume        = 20,
  number        = 3,
  pages         = {335--352},
  doi           = {10.1016/j.neunet.2007.04.016},
  issn          = {0893-6080},
  langid        = {english}
}

@article{kalisman2005neocortical,
  title         = {The neocortical microcircuit as a tabula rasa},
  author        = {Kalisman, Nir and Silberberg, Gilad and Markram, Henry},
  year          = 2005,
  month         = {jan},
  journal       = {Proceedings of the National Academy of Sciences},
  publisher     = {National Acad Sciences},
  volume        = 102,
  number        = 3,
  pages         = {880--885},
  doi           = {10.1073/pnas.0407088102},
  issn          = {0027-8424},
  url           = {http://dx.doi.org/10.1073/pnas.0407088102}
}

@article{kano2018multiple,
  title         = {Multiple phases of climbing fiber synapse elimination in the developing cerebellum},
  author        = {Kano, Masanobu and Watanabe, Takaki and Uesaka, Naofumi and Watanabe, Masahiko},
  year          = 2018,
  month         = {jul},
  journal       = {The Cerebellum},
  publisher     = {Springer},
  volume        = 17,
  pages         = {722--734},
  doi           = {10.1007/s12311-018-0964-z},
  issn          = {1473-4222},
  url           = {http://dx.doi.org/10.1007/s12311-018-0964-z}
}

@article{kao2005contributions,
  title         = {Contributions of an avian basal ganglia--forebrain circuit to real-time modulation of song},
  author        = {Kao, Mimi H. and Doupe, Allison J. and Brainard, Michael S.},
  year          = 2005,
  month         = feb,
  journal       = {Nature},
  publisher     = {Springer Science and Business Media LLC},
  volume        = 433,
  number        = 7026,
  pages         = {638--643},
  doi           = {10.1038/nature03127},
  issn          = {1476-4687},
  url           = {http://dx.doi.org/10.1038/nature03127}
}

@article{kawai2019smallworld,
  title         = {A small-world topology enhances the echo state property and signal propagation in reservoir computing},
  author        = {Kawai, Yuji and Park, Jihoon and Asada, Minoru},
  year          = 2019,
  month         = {apr},
  journal       = {Neural Networks},
  publisher     = {Elsevier},
  volume        = 112,
  pages         = {15--23},
  doi           = {10.1016/j.neunet.2019.01.002},
  issn          = {0893-6080},
  url           = {http://dx.doi.org/10.1016/j.neunet.2019.01.002}
}

@article{konishi1965role,
  title         = {The Role of Auditory Feedback in the Control of Vocalization in the White-Crowned Sparrow},
  author        = {Konishi, Masakazu},
  year          = 1965,
  month         = jan,
  journal       = {Zeitschrift f\"{u}r Tierpsychologie},
  publisher     = {Wiley},
  volume        = 22,
  number        = 7,
  pages         = {770--783},
  doi           = {10.1111/j.1439-0310.1965.tb01688.x},
  issn          = {0044-3573},
  url           = {http://dx.doi.org/10.1111/j.1439-0310.1965.tb01688.x}
}

@article{kremkow2010gating,
  title         = {Gating of signal propagation in spiking neural networks by balanced and correlated excitation and inhibition},
  author        = {Kremkow, Jens and Aertsen, Ad and Kumar, Arvind},
  year          = 2010,
  month         = {nov},
  journal       = {Journal of Neuroscience},
  publisher     = {Society for Neuroscience},
  volume        = 30,
  number        = 47,
  pages         = {15760--15768},
  doi           = {10.1523/jneurosci.3874-10.2010},
  issn          = {0270-6474},
  url           = {http://dx.doi.org/10.1523/jneurosci.3874-10.2010}
}

@article{larkum2013cellular,
  title         = {A Cellular Mechanism for Cortical Associations: An Organizing Principle for the Cerebral Cortex},
  shorttitle    = {A Cellular Mechanism for Cortical Associations},
  author        = {Larkum, Matthew},
  year          = 2013,
  month         = mar,
  journal       = {Trends in Neurosciences},
  volume        = 36,
  number        = 3,
  pages         = {141--151},
  doi           = {10.1016/j.tins.2012.11.006},
  issn          = {0166-2236}
}

@article{lebe2006spontaneous,
  title         = {Spontaneous and evoked synaptic rewiring in the neonatal neocortex},
  author        = {Le B{\'e}, Jean-Vincent and Markram, Henry},
  year          = 2006,
  month         = {aug},
  journal       = {Proceedings of the National Academy of Sciences},
  publisher     = {National Acad Sciences},
  volume        = 103,
  number        = 35,
  pages         = {13214--13219},
  doi           = {10.1073/pnas.0604691103},
  issn          = {0027-8424},
  url           = {http://dx.doi.org/10.1073/pnas.0604691103}
}

@article{leonardo2005ensemble,
  title         = {Ensemble coding of vocal control in birdsong},
  author        = {Leonardo, Anthony and Fee, Michale S},
  year          = 2005,
  month         = {jan},
  journal       = {Journal of Neuroscience},
  publisher     = {Soc Neuroscience},
  volume        = 25,
  number        = 3,
  pages         = {652--661},
  doi           = {10.1523/jneurosci.3036-04.2005},
  issn          = {0270-6474},
  url           = {http://dx.doi.org/10.1523/jneurosci.3036-04.2005}
}

@article{london2005dendritic,
  title         = {Dendritic computation},
  author        = {London, Michael and H{\"a}usser, Michael},
  year          = 2005,
  month         = {jul},
  journal       = {Annual Review of Neuroscience},
  publisher     = {Annual Reviews},
  volume        = 28,
  pages         = {503--532},
  doi           = {10.1146/annurev.neuro.28.061604.135703},
  issn          = {0147-006X},
  url           = {http://dx.doi.org/10.1146/annurev.neuro.28.061604.135703}
}

@article{lukosevicius2009reservoir,
  title         = {Reservoir Computing Approaches to Recurrent Neural Network Training},
  author        = {Luko{\v s}evi{\v c}ius, Mantas and Jaeger, Herbert},
  year          = 2009,
  month         = aug,
  journal       = {Computer Science Review},
  volume        = 3,
  number        = 3,
  pages         = {127--149},
  doi           = {10.1016/j.cosrev.2009.03.005},
  issn          = {1574-0137}
}

@article{lynch2016rhythmic,
  title         = {Rhythmic continuous-time coding in the songbird analog of vocal motor cortex},
  author        = {Lynch, Galen F and Okubo, Tatsuo S and Hanuschkin, Alexander and Hahnloser, Richard HR and Fee, Michale S},
  year          = 2016,
  month         = {may},
  journal       = {Neuron},
  publisher     = {Elsevier},
  volume        = 90,
  number        = 4,
  pages         = {877--892},
  doi           = {10.1016/j.neuron.2016.04.021},
  issn          = {0896-6273},
  url           = {http://dx.doi.org/10.1016/j.neuron.2016.04.021}
}

@article{maass2002realtime,
  title         = {Real-time computing without stable states: A new framework for neural computation based on perturbations},
  author        = {Maass, Wolfgang and Natschl{\"a}ger, Thomas and Markram, Henry},
  year          = 2002,
  month         = {nov},
  journal       = {Neural Computation},
  publisher     = {MIT Press},
  volume        = 14,
  number        = 11,
  pages         = {2531--2560},
  doi           = {10.1162/089976602760407955},
  issn          = {0899-7667},
  url           = {http://dx.doi.org/10.1162/089976602760407955}
}

@article{maes2020learning,
  title         = {Learning spatiotemporal signals using a recurrent spiking network that discretizes time},
  author        = {Maes, Amadeus and Barahona, Mauricio and Clopath, Claudia},
  year          = 2020,
  month         = jan,
  journal       = {PLOS Computational Biology},
  publisher     = {Public Library of Science (PLoS)},
  volume        = 16,
  number        = 1,
  pages         = {1--26},
  doi           = {10.1371/journal.pcbi.1007606},
  issn          = {1553-7358},
  url           = {http://dx.doi.org/10.1371/journal.pcbi.1007606},
  editor        = {Richards, Blake A.}
}

@article{malvache2016awake,
  title         = {Awake hippocampal reactivations project onto orthogonal neuronal assemblies},
  author        = {Malvache, Arnaud and Reichinnek, Susanne and Villette, Vincent and Haimerl, Caroline and Cossart, Rosa},
  year          = 2016,
  month         = {sep},
  journal       = {Science},
  publisher     = {American Association for the Advancement of Science},
  volume        = 353,
  number        = 6305,
  pages         = {1280--1283},
  doi           = {10.1126/science.aaf3319},
  issn          = {0036-8075},
  url           = {http://dx.doi.org/10.1126/science.aaf3319}
}

@article{murray2019local,
  title         = {Local online learning in recurrent networks with random feedback},
  author        = {Murray, James M.},
  year          = 2019,
  month         = may,
  journal       = {eLife},
  publisher     = {eLife Sciences Publications, Ltd},
  volume        = 8,
  doi           = {10.7554/elife.43299},
  issn          = {2050-084X},
  url           = {http://dx.doi.org/10.7554/eLife.43299}
}

@article{nicola2017supervised,
  title         = {Supervised learning in spiking neural networks with FORCE training},
  author        = {Nicola, Wilten and Clopath, Claudia},
  year          = 2017,
  month         = {dec},
  journal       = {Nature Communications},
  publisher     = {Nature Publishing Group UK London},
  volume        = 8,
  number        = 1,
  pages         = 2208,
  doi           = {10.1038/s41467-017-01827-3},
  issn          = {2041-1723},
  url           = {http://dx.doi.org/10.1038/s41467-017-01827-3}
}

@article{payeur2019classes,
  title         = {Classes of Dendritic Information Processing},
  author        = {Payeur, Alexandre and B{\'e}{\"i}que, Jean-Claude and Naud, Richard},
  year          = 2019,
  month         = oct,
  journal       = {Current Opinion in Neurobiology},
  series        = {Computational {{Neuroscience}}},
  volume        = 58,
  pages         = {78--85},
  doi           = {10.1016/j.conb.2019.07.006},
  issn          = {0959-4388},
  langid        = {english}
}

@article{peters2014emergence,
  title         = {Emergence of reproducible spatiotemporal activity during motor learning},
  author        = {Peters, Andrew J and Chen, Simon X and Komiyama, Takaki},
  year          = 2014,
  month         = {may},
  journal       = {Nature},
  publisher     = {Nature Publishing Group UK London},
  volume        = 510,
  number        = 7504,
  pages         = {263--267},
  doi           = {10.1038/nature13235},
  issn          = {0028-0836},
  url           = {http://dx.doi.org/10.1038/nature13235}
}

@article{petrovici2014characterization,
  title         = {Characterization and {{Compensation}} of {{Network-Level Anomalies}} in {{Mixed-Signal Neuromorphic Modeling Platforms}}},
  author        = {Petrovici, Mihai A. and Vogginger, Bernhard and M{\"u}ller, Paul and Breitwieser, Oliver and Lundqvist, Mikael and Muller, Lyle and Ehrlich, Matthias and Destexhe, Alain and Lansner, Anders and Sch{\"u}ffny, Ren{\'e} and Schemmel, Johannes and Meier, Karlheinz},
  year          = 2014,
  month         = oct,
  journal       = {PLOS ONE},
  publisher     = {Public Library of Science},
  volume        = 9,
  number        = 10,
  pages         = {e108590},
  doi           = {10.1371/journal.pone.0108590},
  issn          = {1932-6203},
  langid        = {english}
}

@article{poirazi2020illuminating,
  title         = {Illuminating dendritic function with computational models},
  author        = {Poirazi, Panayiota and Papoutsi, Athanasia},
  year          = 2020,
  month         = {may},
  journal       = {Nature Reviews Neuroscience},
  publisher     = {Nature Publishing Group UK London},
  volume        = 21,
  number        = 6,
  pages         = {303--321},
  doi           = {10.1038/s41583-020-0301-7},
  issn          = {1471-003X},
  url           = {http://dx.doi.org/10.1038/s41583-020-0301-7}
}

@inproceedings{sacramento2018dendritic,
  title         = {Dendritic Cortical Microcircuits Approximate the Backpropagation Algorithm},
  author        = {Sacramento, Jo{\~a}o and Ponte Costa, Rui and Bengio, Yoshua and Senn, Walter},
  year          = 2018,
  booktitle     = {Advances in {{Neural Information Processing Systems}}},
  publisher     = {Curran Associates, Inc.},
  volume        = 31,
  url           = {https://proceedings.neurips.cc/paper_files/paper/2018/file/1dc3a89d0d440ba31729b0ba74b93a33-Paper.pdf},
  editor        = {S. Bengio and H. Wallach and H. Larochelle and K. Grauman and N. Cesa-Bianchi and R. Garnett}
}

@article{shadlen1994noise,
  title         = {Noise, neural codes and cortical organization},
  author        = {Shadlen, Michael N and Newsome, William T},
  year          = 1994,
  month         = {aug},
  journal       = {Current Opinion in Neurobiology},
  publisher     = {Elsevier},
  volume        = 4,
  number        = 4,
  pages         = {569--579},
  doi           = {10.1016/0959-4388(94)90059-0},
  issn          = {0959-4388},
  url           = {http://dx.doi.org/10.1016/0959-4388(94)90059-0}
}

@article{shu2003turning,
  title         = {Turning on and off recurrent balanced cortical activity},
  author        = {Shu, Yousheng and Hasenstaub, Andrea and McCormick, David A},
  year          = 2003,
  month         = {may},
  journal       = {Nature},
  publisher     = {Nature Publishing Group},
  volume        = 423,
  number        = 6937,
  pages         = {288--293},
  doi           = {10.1038/nature01616},
  issn          = {0028-0836},
  url           = {http://dx.doi.org/10.1038/nature01616}
}

@article{sreenivasan2019delay,
  title         = {The what, where and how of delay activity},
  author        = {Sreenivasan, Kartik K and D'Esposito, Mark},
  year          = 2019,
  month         = {may},
  journal       = {Nature Reviews Neuroscience},
  publisher     = {Nature Publishing Group UK London},
  volume        = 20,
  number        = 8,
  pages         = {466--481},
  doi           = {10.1038/s41583-019-0176-7},
  issn          = {1471-003X},
  url           = {http://dx.doi.org/10.1038/s41583-019-0176-7}
}

@book{stuart2016dendrites,
  title         = {Dendrites},
  author        = {Stuart, Greg and Spruston, Nelson and H{\"a}usser, Michael},
  year          = 2016,
  month         = {mar},
  publisher     = {Oxford University Press},
  doi           = {10.1093/acprof:oso/9780198745273.001.0001},
  url           = {http://dx.doi.org/10.1093/acprof:oso/9780198745273.001.0001}
}

@article{sussillo2009generating,
  title         = {Generating {{Coherent Patterns}} of {{Activity}} from {{Chaotic Neural Networks}}},
  author        = {Sussillo, David and Abbott, Larry F.},
  year          = 2009,
  month         = aug,
  journal       = {Neuron},
  publisher     = {Elsevier},
  volume        = 63,
  number        = 4,
  pages         = {544--557},
  doi           = {10.1016/j.neuron.2009.07.018},
  issn          = {0896-6273},
  langid        = {english},
  pmid          = 19709635
}

@article{tanaka2019recent,
  title         = {Recent advances in physical reservoir computing: A review},
  author        = {Tanaka, Gouhei and Yamane, Toshiyuki and H{\'e}roux, Jean Benoit and Nakane, Ryosho and Kanazawa, Naoki and Takeda, Seiji and Numata, Hidetoshi and Nakano, Daiju and Hirose, Akira},
  year          = 2019,
  month         = {jul},
  journal       = {Neural Networks},
  publisher     = {Elsevier},
  volume        = 115,
  pages         = {100--123},
  doi           = {10.1016/j.neunet.2019.03.005},
  issn          = {0893-6080},
  url           = {http://dx.doi.org/10.1016/j.neunet.2019.03.005}
}

@article{urbanczik2014learning,
  title         = {Learning by the {{Dendritic Prediction}} of {{Somatic Spiking}}},
  author        = {Urbanczik, Robert and Senn, Walter},
  year          = 2014,
  month         = feb,
  journal       = {Neuron},
  volume        = 81,
  number        = 3,
  pages         = {521--528},
  doi           = {10.1016/j.neuron.2013.11.030},
  issn          = {08966273},
  langid        = {english}
}

@article{vanvreeswijk1996chaos,
  title         = {Chaos in {{Neuronal Networks}} with {{Balanced Excitatory}} and {{Inhibitory Activity}}},
  author        = {{van Vreeswijk}, C. and Sompolinsky, H.},
  year          = 1996,
  month         = dec,
  journal       = {Science},
  publisher     = {American Association for the Advancement of Science},
  volume        = 274,
  number        = 5293,
  pages         = {1724--1726},
  doi           = {10.1126/science.274.5293.1724}
}

@article{wehr2003balanced,
  title         = {Balanced inhibition underlies tuning and sharpens spike timing in auditory cortex},
  author        = {Wehr, Michael and Zador, Anthony M.},
  year          = 2003,
  month         = nov,
  journal       = {Nature},
  publisher     = {Nature Publishing Group},
  volume        = 426,
  number        = 6965,
  pages         = {442--446},
  doi           = {10.1038/nature02116},
  issn          = {1476-4687}
}

@article{werbos1990backpropagation,
  title         = {Backpropagation through time: what it does and how to do it},
  author        = {Werbos, Paul John},
  year          = 1990,
  month         = oct,
  journal       = {Proceedings of the IEEE},
  publisher     = {Institute of Electrical and Electronics Engineers (IEEE)},
  volume        = 78,
  number        = 10,
  pages         = {1550--1560},
  doi           = {10.1109/5.58337},
  issn          = {1558-2256}
}

@article{zador2019critique,
  title         = {A Critique of Pure Learning and What Artificial Neural Networks Can Learn from Animal Brains},
  author        = {Zador, Anthony M.},
  year          = 2019,
  month         = aug,
  journal       = {Nature Communications},
  publisher     = {Nature Publishing Group},
  volume        = 10,
  number        = 1,
  pages         = 3770,
  doi           = {10.1038/s41467-019-11786-6},
  issn          = {2041-1723},
  langid        = {english}
}

@article{zipser1989subgrouping,
  title         = {A subgrouping strategy that reduces complexity and speeds up learning in recurrent networks},
  author        = {Zipser, David},
  year          = 1989,
  month         = {dec},
  journal       = {Neural Computation},
  publisher     = {MIT Press One Rogers Street, Cambridge},
  volume        = 1,
  number        = 4,
  pages         = {552--558},
  doi           = {10.1162/neco.1989.1.4.552},
  issn          = {0899-7667},
  url           = {http://dx.doi.org/10.1162/neco.1989.1.4.552}
}

@article{zylberberg2017mechanisms,
  title         = {Mechanisms of persistent activity in cortical circuits: possible neural substrates for working memory},
  author        = {Zylberberg, Joel and Strowbridge, Ben W},
  year          = 2017,
  month         = {jul},
  journal       = {Annual Review of Neuroscience},
  publisher     = {Annual Reviews},
  volume        = 40,
  number        = 1,
  pages         = {603--627},
  doi           = {10.1146/annurev-neuro-070815-014006},
  issn          = {0147-006X},
  url           = {http://dx.doi.org/10.1146/annurev-neuro-070815-014006}
}
